\documentclass[twocolumn]{IEEEtran}
\usepackage{amsmath}
\usepackage{amssymb}
\usepackage{graphics}
\usepackage{graphicx}

\usepackage{fixltx2e}
\usepackage{epstopdf}
\usepackage{amsmath}
\usepackage{tikz}
\usetikzlibrary{shapes.geometric,shapes, shapes.misc, patterns, decorations.text,plotmarks,3d}
\usepackage{pgfplots}
\pgfplotsset{compat=newest} 
\pgfplotsset{plot coordinates/math parser=false} 
\newlength\figureheight 
\newlength\figurewidth 
\usepackage{bm}
\usepackage{cite}
\usepackage[capitalize]{cleveref}
\crefname{algocf}{Algorithm}{Algorithms}
\crefname{AlgoLine}{Line}{Algorithms}
\crefname{equation}{\unskip}{\unskip}
\usepackage{comment}
\usepackage{euscript}
\usepackage[lined,boxed,commentsnumbered,linesnumbered, ruled]{algorithm2e}
\usepackage{algorithmic}

\usepackage{tikz}
\usetikzlibrary{shapes.geometric, shapes, patterns}
\usepackage{adjustbox}

\newif\ifonecolumn
\ifCLASSINFOpdf
\else
\fi

\makeatletter
\newcommand*{\rom}[1]{\expandafter\@slowromancap\romannumeral #1@}
\makeatother

\usetikzlibrary{external}
\newcommand{\cA}{{\mathsf{A}}}
\newcommand{\cB}{{\mathsf{B}}}
\newcommand{\nA}{n_{\mathsf{A}}}
\newcommand{\nB}{n_{\mathsf{B}}}

\newcommand{\D}{\boldsymbol{D}}
\newcommand{\A}{\boldsymbol{A}}

\renewcommand{\c}[2]{\ensuremath{\mathcal {#1}_{#2}}}		% Set #1--> Set #2-->subscript
\newcommand\scalemath[2]{\scalebox{#1}{\mbox{\ensuremath{\displaystyle #2}}}}
\DeclareMathOperator*{\argmin}{arg\,min}
\DeclareMathOperator*{\readd}{read}

\newtheorem{theorem}{Theorem}

\newtheorem{lemma}{Lemma}
\newtheorem{definition}{Definition}
\newtheorem{proposition}{Proposition}
\newtheorem{example}{Example}
\newtheorem{remark}{Remark}

\begin{document}
\title{Code Constructions  for Distributed Storage With Low Repair Bandwidth and Low Repair Complexity}
\author{Siddhartha~Kumar,~\IEEEmembership{Student~Member,~IEEE}, Alexandre~Graell~i~Amat,~\IEEEmembership{Senior~Member,~IEEE},       Iryna~Andriyanova,~\IEEEmembership{Member,~IEEE}, Fredrik~Br\"{a}nnstr\"{o}m,~\IEEEmembership{Member,~IEEE}, and~Eirik~Rosnes,~\IEEEmembership{Senior~Member,~IEEE}% <-this % stops a space
\thanks{Parts of this paper were presented at the IEEE Global Communications Conference (GLOBECOM), San Diego, CA, December 2015.  %} % <-this % stops a space
This work was partially funded by the Research Council of Norway (grant 240985/F20), Simula@UiB, and the Swedish Research Council (grant \#2016-04253).}%
\thanks{S.\ Kumar was with the Department of Electrical Engineering, Chalmers University of Technology, SE-41296 Gothenburg, Sweden. He is now with Simula@UiB, N-5020 Bergen, Norway (e-mail: kumarsi@simula.no).}
\thanks{A.\ Graell i Amat  and F. Br\"{a}nnstr\"{o}m are with the Department of Electrical Engineering, Chalmers University of Technology, SE-41296 Gothenburg, Sweden (e-mail: \{alexandre.graell, fredrik.brannstrom\}@chalmers.se).}
\thanks{I.\ Andriyanova is with the ETIS-UMR8051 group, ENSEA/University of Cergy-Pontoise/CNRS, 95015 Cergy, France (e-mail: iryna.andriyanova@ensea.fr).}% <-this % stops a space
\thanks{E.\ Rosnes is with Simula@UiB, N-5020 Bergen, Norway (e-mail: eirikrosnes@simula.no).}
}

% make the title area
\maketitle

% As a general rule, do not put math, special symbols or citations
% in the abstract or keywords.
\begin{abstract}
We present the construction of a family of erasure correcting codes for distributed storage  that achieve low repair bandwidth and complexity at the expense of a lower fault tolerance. The construction is based on two classes of codes, where the primary goal of the first class of codes is to provide fault tolerance, while the second class  aims at reducing the repair bandwidth and  repair complexity.  The repair procedure is a two-step procedure where parts of the failed node are repaired in the first step using the first code. The downloaded symbols during the first step are cached in the memory and used to repair the remaining erased data symbols at minimal additional read cost during the second step. The first class of codes is based on MDS codes modified using piggybacks, while the second class  is designed to reduce the number of additional symbols that need to be downloaded to repair the remaining erased symbols.  We numerically show that the proposed codes achieve better repair bandwidth compared to MDS codes, codes constructed using piggybacks, and local reconstruction/Pyramid codes,  while a better repair complexity is achieved when compared to MDS, Zigzag, Pyramid codes, and codes constructed using piggybacks.
\end{abstract}
%\begin{abstract}
%We present the construction of a family of erasure correcting codes for distributed storage systems that achieve low repair bandwidth and low repair complexity. The construction is based on two classes of codes, where the primary goal of the first class of codes \textcolor{blue}{is to provide fault tolerance}, while the second class of codes aims at reducing the repair bandwidth and the repair complexity.  The repair procedure is a two-step procedure where parts of the failed node are repaired in the first step using the first code. The downloaded symbols during the first step are cached in the memory and used to repair the remaining erased data symbols at no additional read cost during the second step of the repair process. The first class of codes is based on MDS codes modified using piggybacks, while the second class of codes is designed to reduce the number of additional symbols that need to be downloaded to repair the remaining erased symbols.  We show that the proposed codes achieve better repair bandwidth compared to MDS, Piggyback, and local reconstruction codes,  while a better repair complexity is achieved when compared to MDS, Zigzag, and Piggyback codes.
%\end{abstract}

% For peer review papers, you can put extra information on the cover
% page as needed:
% \ifCLASSOPTIONpeerreview
% \begin{center} \bfseries EDICS Category: 3-BBND \end{center}
% \fi
%
% For peerreview papers, this IEEEtran command inserts a page break and
% creates the second title. It will be ignored for other modes.
\IEEEpeerreviewmaketitle
\begin{IEEEkeywords}
Code concatenation, codes for distributed storage, piggybacking, repair bandwidth, repair complexity.
\end{IEEEkeywords}

\section{Introduction}

% The very first letter is a 2 line initial drop letter followed
% by the rest of the first word in caps.
% 
% form to use if the first word consists of a single letter:
% \IEEEPARstart{A}{demo} file is ....
% 
% form to use if you need the single drop letter followed by
% normal text (unknown if ever used by the IEEE):
% \IEEEPARstart{A}{}demo file is ....
% 
% Some journals put the first two words in caps:
% \IEEEPARstart{T}{his demo} file is ....
% 
% Here we have the typical use of a "T" for an initial drop letter
% and "HIS" in caps to complete the first word.

\IEEEPARstart{I}n recent years, there has been a widespread adoption of distributed storage systems (DSSs) as a viable storage technology for Big Data. Distributed storage provides an inexpensive storage solution for storing large amounts of data. Formally, a DSS is a network of numerous inexpensive disks (or nodes) where data is stored in a distributed fashion. Storage nodes are prone to failures, and thus to losing the stored data. Reliability against node failures (commonly referred to as fault tolerance) is achieved by means of erasure correcting codes (ECCs).  ECCs are a way of introducing structured redundancy, and for a DSS, it means addition of redundant nodes. In case of a node failure, these redundant nodes allow complete recovery of the data stored. Since ECCs have a limited fault tolerance, to maintain the initial state of reliability, when a node fails a new node needs to be added to the DSS network and populated with data. The problem of repairing a failed node is known as the repair problem.

Current DSSs like Google File System \rom{2} and Quick File System use a family of Reed-Solomon (RS)  ECCs \cite{hua14}. Such codes come under a broader family of maximum distance separable (MDS) codes. MDS codes are optimal in terms of the fault tolerance/storage overhead tradeoff. However, the repair of a failed node requires the retrieval of large amounts of data from a large subset of nodes. Therefore, in the recent years, the design of ECCs that reduce the cost of repair has attracted significant attention. Pyramid codes \cite{hua07} were one of the first code constructions that addressed this problem. In particular, Pyramid codes are a class of non-MDS codes that aim at reducing the number of nodes that need to be contacted to repair a single failed node, known as the repair locality. Other non-MDS codes that reduce the repair locality are local reconstruction codes (LRCs) \cite{hua12} and locally repairable codes \cite{sat13,Pap14}. Such codes achieve a low repair locality by ensuring that the parity symbols are a function of a small number of data symbols, which also entails a low repair complexity, defined as the  number of elementary additions required to repair a failed node. Furthermore, for a fixed locality LRCs and Pyramid codes achieve the optimal fault tolerance.

Another important parameter related to the repair is the repair bandwidth, defined as the number of symbols downloaded to repair a single failed node. Dimakis \emph{et al.} \cite{dim10} derived an optimal repair bandwidth-storage per node tradeoff curve and defined two new classes of codes for DSSs known as minimum storage regenerating (MSR) codes and minimum bandwidth regenerating (MBR) codes that are at the two extremal points of the tradeoff curve. 
MSR codes are MDS codes with the best storage efficiency, i.e., they require a minimum storage of data per node (referred to as the sub-packetization level). On the other hand, MBR codes achieve the minimum repair bandwidth.
%MSR codes are MDS codes that have the lowest possible repair bandwidth while storing the minimum amount of data per node (referred to as the sub-packetization level). On the other hand, MBR codes achieve the lowest repair bandwidth while having the maximum sub-packetization level.
 Product-Matrix MBR (PM-MBR) codes and Fractional Repetition (FR) codes in \cite{Ras11} and \cite{Rou10}, respectively, are examples of MBR codes. In particular, FR codes achieve low repair complexity at the cost of high storage overheads.
 Codes such as minimum disk input/output repairable (MDR) codes \cite{wan14} and Zigzag codes \cite{tam13} strictly fall under the class of MSR codes. These codes have a high sub-packetization level. Alternatively, the MSR codes presented in \cite{Cad11, Aga15, Li15, Wan16, Sas16, Ye17, Li17, Rav17} achieve the minimum possible sub-packetization level. 
 
Piggyback codes presented in \cite{Ras17} are another class of codes that achieve a sub-optimal reduction in repair bandwidth with a much lower sub-packetization level in comparison to MSR codes, using the concept of \emph{piggybacking}. Piggybacking consists of adding carefully chosen linear combinations of data symbols (called piggybacks) to the parity symbols of a given ECC. This results in a lower repair bandwidth at the expense of a higher complexity in encoding and repair operations.
More recently, the authors in \cite{Hou16} presented a family of codes that reduce the encoding and repair complexity of PM-MBR codes while maintaining the same level of fault tolerance and repair bandwidth. However, this comes at the cost of large alphabet size. In \cite{Hou17}, binary MDS array codes that achieve optimal repair bandwidth and low repair complexity were introduced, with the caveat that the file size is asymptotic and that the fault tolerance is limited to $3$.

In this paper, we propose a family of non-MDS ECCs that achieve low repair bandwidth and low repair complexity while keeping the field size relatively small and having variable fault tolerance. In particular, we propose a systematic code construction based on two classes of parity symbols. Correspondingly, there are two classes of parity nodes. The first class of parity nodes, whose primary goal is to provide erasure correcting capability, is constructed using an MDS code modified by applying specially designed piggybacks to some of its code symbols.
As a secondary goal, the first class of parity nodes enable to repair a number of data symbols at a low repair cost by downloading piggybacked symbols.
 The second class of parity nodes is constructed using a block code whose parity symbols are obtained through simple additions. The purpose of this class of parity nodes is not to enhance the erasure correcting capability, but rather to facilitate node repair at low repair bandwidth and low repair complexity by repairing the remaining failed symbols in the node. %We compare the proposed codes with MDR codes, Zigzag codes, Piggyback codes, and LRCs \cite{hua12}, both in terms of repair bandwidth and repair complexity.
Compared to \cite{kum15a}, we provide two constructions for the second class of parity nodes. The first one is given by a simple equation that represents the algorithmic construction in \cite{kum15a}. The second one is a heuristic construction that is more involved, but further reduces the repair bandwidth in some cases. Furthermore, we provide explicit formulas for the fault tolerance, repair bandwidth, and repair complexity of the proposed codes and numerically compare with other codes in the literature.
%We provide two constructions for the second class of parity nodes. The first one is given by a simple equation. The second one is a heuristic construction which is more involved but further improves the repair bandwidth in some cases. 
The proposed codes achieve better repair bandwidth compared to MDS codes, Piggyback codes, generalized Piggyback codes \cite{Yua16}, and exact-repairable MDS codes \cite{Raw17}. For certain code parameters, we also see that the proposed codes have better repair bandwidth compared to LRCs and Pyramid codes. Furthermore, they achieve better repair complexity than Zigzag codes, MDS codes, Piggyback codes, generalized Piggyback codes, exact-repairable MDS codes, and binary addition and shift implementable cyclic-convolutional (BASIC) PM-MBR codes \cite{Hou16}. Also, for certain code parameters, the codes have better repair complexity than Pyramid codes.
The improvements over MDS codes, MSR codes, and the classes of Piggyback codes come at the expense of a lower fault tolerance in general.

\section{System Model and Code Construction} \label{sec:system_model}

\begin{figure}[!t]
\centering
\includegraphics{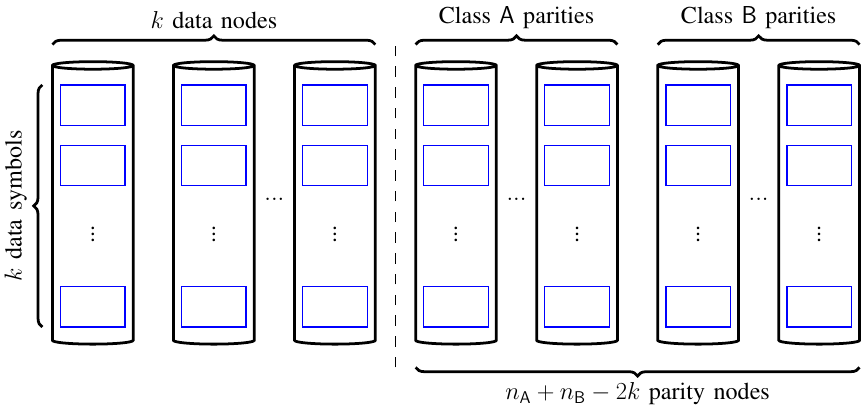}
\vspace{-3ex}
\caption{System model of the DSS.}
\label{Fig:SystemModel}
\vspace{-3ex}
\end{figure}

We consider the DSS depicted in Fig.~\ref{Fig:SystemModel}, consisting of storage nodes, of which $k$ are data nodes and $n-k$ are parity nodes. Consider a file that needs to be stored on the DSS. We represent a file as a $k\times k$ matrix $\bm D=[d_{i,j}]$, called the data array, over  $\text{GF}(q)$, where $\text{GF}(q)$ denotes the Galois field of size $q$, with $q$ being a prime number or a power of a prime number. In order to achieve reliability against node failures, the matrix $\bm D$ is encoded using an $(n,k)$ vector code \cite{Bla95} to obtain a code matrix $\bm C=[c_{i,j}]$, referred to as the code array, of size $k\times n$, $c_{i,j}\in\text{GF}(q)$. The symbol $c_{i,j}$ in $\bm C$ is then stored at the $i$-th row of the $j$-th node in the DSS. Thus, each node stores $k$ symbols.
%a vector $\bm d=(d_{0,0},d_{0,1},\ldots,d_{0,k-1},\ldots,d_{k-1,k-1})$ having message symbols $d_{i,j}\in\text{GF}(q^p)$, where $\text{GF}(q^p)$ denotes the finite field of size $q^p$ where $q$ is a prime and $p$ is a positive integer. In order to achieve reliability against node failures, such a vector, $\bm d$, of length $k^2$ is encoded using an $(nk,k^2)$ code to obtain a code vector $\bm c=(c_{0,0},c_{0,1},\ldots,c_{0,n-1},\ldots,c_{k-1,n-1})$ of length $nk$ with each code symbol $c_{i,j}\in\text{GF}(q^p)$. With some abuse of notation, in the following, we refer to these codes as $(n,k)$ ECCs. The code symbols thus obtained are arranged in a code array $\bm C=[c_{i,j}]$ of size $k\times n$, where the symbols in the $j$-th column are stored in the $j$-th node. 
Each row in $\bm C$ is referred to as a stripe so that each file in the DSS is stored over $k$ stripes in $n$ storage nodes. We consider the $(n,k)$ code to be systematic, which means that $c_{i,j}=d_{i,j}$ for $i,j=0,\ldots,k-1$. Correspondingly, we refer to the $k$ nodes storing systematic symbols as data nodes and the remaining $n-k$ nodes containing parity symbols only as parity nodes. The efficiency of the code is determined by the code rate, given by $R=k^2/kn=k/n$. Alternatively, the inverse of the code rate is referred to as the storage overhead.

For later use, we denote the set of message symbols in the $k$ data nodes as $\mathcal D=\{d_{i,j}\}$ and by $\mathcal P_t$, $t=k,\ldots,n-1$, the set of parity symbols in the $t$-th node. Subsequently, we define the set $\mathcal D_{\mathcal I}\subseteq\mathcal D$ as 
\begin{align*}
	\mathcal D_{\mathcal I}=\{d_{i,j}\in\mathcal D \mid (i,j)\in\mathcal I\},
\end{align*}
where $\mathcal I$ is an arbitrary index set. We also define the operator $(a+b)_k \triangleq (a+b) \bmod k$ for integers $a$ and $b$.

Our main goal is to construct codes that yield low repair bandwidth and low repair complexity of a single failed data node. We focus on the repair of data nodes since the raw data is stored on these nodes and the users can readily access the data through these nodes. Thus, their survival is a crucial aspect of a DSS.
To this purpose, we construct a family of systematic $(n,k)$ codes consisting of two different classes of parity symbols. Correspondingly, there are two classes of parity nodes, referred to as Class $\cA$ and Class $\cB$ parity nodes, as shown in Fig.~\ref{Fig:SystemModel}. Class $\cA$ and Class $\cB$ parity nodes are built using an $(\nA,k)$ code and an $(\nB, k)$ code, respectively, such that $n=\nA+\nB-k$. In other words, the parity nodes from the $(n,k)$ code
%\footnote{With some abuse of language we refer to the nodes storing the parity symbols of a code as the parity nodes of the code.}
 correspond to the parity nodes of Class $\cA$ and Class $\cB$ codes. The primary goal of Class $\cA$ parity nodes is to achieve a good erasure correcting capability, while the purpose of Class $\cB$ nodes is to yield low repair bandwidth and low repair complexity. In particular, we focus on the repair of data nodes. The repair bandwidth (in bits) per node, denoted by $\gamma$, is proportional to the average number of symbols (data and parity) that need to be downloaded to repair a data symbol, denoted by $\lambda$. More precisely, let $\beta$ be the sub-packetization level of the DSS, which is the number of symbols per node.\footnote{For our code construction, $\beta=k$, but this is not the case in general.} Then,
\begin{align}
\lambda = \frac{\gamma}{\nu \beta},%=\frac{\gamma}{\nu k},
\label{eq:lambda}
\end{align}
where $\nu = m \lceil \log_2 p \rceil$ is the size (in bits) of a symbol in $\text{GF}(q)$, where $q=p^m$ for some prime number $p$ and positive integer $m$. $\lambda$ can be interpreted as the repair bandwidth normalized by the size (in bits) of a node, and will be referred to as the \emph{normalized}  repair bandwidth. %Therefore, in the rest of the paper, we will use $\lambda$ to refer to the normalized repair bandwidth.

The main principle behind our code construction is the following. The repair is performed one symbol at a time. After the repair of a data symbol is accomplished, the symbols read to repair that symbol are cached in the memory. Therefore, they can be used to repair the remaining data symbols at no additional read cost. The proposed codes are constructed in such a way that the repair of a new data symbol requires a low additional read cost (defined as the number of additional symbols that need to be read to repair the data symbol), so that $\lambda$ (and hence $\gamma$) is kept low. %Since we will often use the concepts of read cost and additional read cost in the remainder of the paper, we define them in the following.
\begin{definition}
The \emph{read cost} of a symbol is the number of symbols that need to be read to repair the symbol. For a symbol that is repaired after some others, the \emph{additional read cost} is defined as the number of additional symbols that need to be read to repair the symbol. (Note that symbols previously read to repair other data symbols are already cached in the memory and to repair a new symbol only some extra symbols may need to be read.)
\end{definition}

\section{Class $\cA$ Parity Nodes} \label{Sec:ClassA}

Class $\cA$ parity nodes are constructed using a modified $(\nA, k)$ MDS code, with $k+ 2 \le \nA < 2k$, over $\text{GF}(q)$. In particular, we start from an $(\nA, k)$ MDS code and apply piggybacks \cite{Ras17} to some of the parity symbols. The construction of Class $\cA$ parity nodes is performed in two steps as follows.
\begin{enumerate}
\item[1)] Encode each row of the data array using an $(\nA,k)$ MDS code (same code for each row). The parity symbol $p^{\cA}_{i,j}$ is obtained  as\footnote{We use the superscript $\cA$ to indicate that the parity symbol is stored in a Class $\cA$ parity node.}
\begin{align}
\label{eq:pij}
p^{\cA}_{i,j}=\sum_{l=0}^{k-1} \alpha_{l,j}d_{i,l},~j=k,\ldots ,\nA-1,
\end{align}
where $\alpha_{l,j}$ denotes a coefficient in $\text{GF}(q)$ and $i=0,\ldots,k-1$.
Store the parity symbol in the corresponding row of the code array. Overall, $k(\nA-k)$ parity symbols are generated.
\item[2)] Modify some of the parity symbols by adding piggybacks. Let $\tau$, $1\leq\tau\leq \nA-k-1$, be the number of piggybacks introduced per row. The parity symbol $p_{i,u}^{\cA}$ is updated as
\begin{align}
\label{eq:piu}
p^{\cA,\mathsf p}_{i,u}= p^{\cA}_{i,u}  + d_{(i+u-\nA+\tau+1)_k, i},
\end{align}
where $ u=\nA-\tau,\ldots,\nA-1$, the second term in the summation is the piggyback, and the superscript $\mathsf p$ in $p^{\cA,\mathsf p}_{i,u}$ indicates that the parity symbol contains piggybacks.
\end{enumerate}

The fault tolerance (i.e., the number of node failures that can be tolerated) of Class $\cA$ codes is given in the following theorem.

\begin{theorem} \label{th:fault_tol}
An $(\nA,k)$ Class $\cA$ code with $\tau$ piggybacks per row can tolerate
\begin{align*}
	f=\begin{cases}
	 \nA-k-\tau+\bigg\lfloor\frac{\sqrt{(\nA-k-\tau)^2+4k}-(\nA-k-\tau)}{2}\bigg\rfloor & \text{if}\,\, \tau\geq\xi\\
	 \nA-k & \text{if}\,\, \tau<\xi	
	\end{cases}
\end{align*}
node failures, where $\xi=\frac{\sqrt{(\nA-k-\tau)^2+4k}-(\nA-k-\tau)}{2}$.
%a minimum of $\nA-k-\tau+1$ node failures.
\label{th:ECC}
\end{theorem}
	\begin{IEEEproof}
		See Appendix~\ref{App: Proof_of_fault_tolerance}.
	\end{IEEEproof}
	
%\cref{th:ECC} presents a lower bound on the fault tolerance of the Class $\cA$ code. 
We remark that for $\tau<\xi$, Class $\cA$ codes are MDS codes.

\begin{figure}[t]
		\centering
\includegraphics{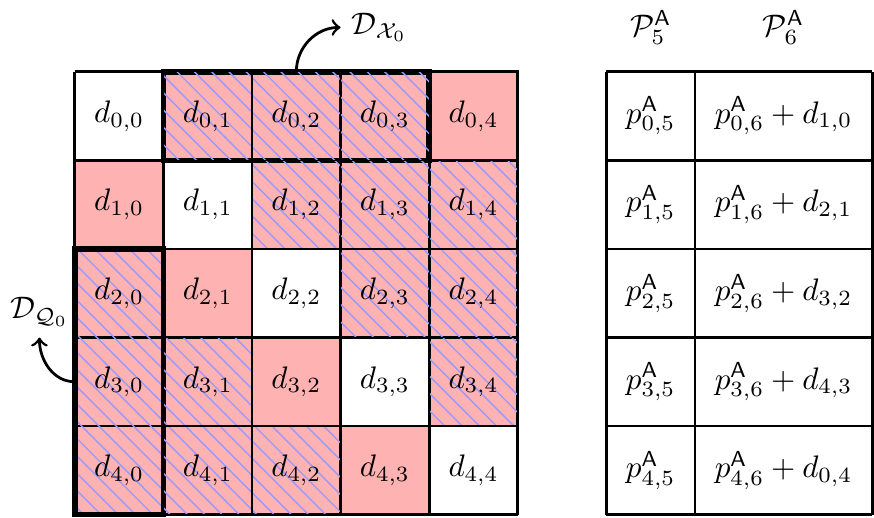}
		\vspace{-2ex}
		\caption{A $(7, 5)$ Class $\cA$ code with $\tau=1$ constructed from a $(7, 5)$ MDS code. %$N_0, N_1, ...$ denote the data nodes while 
		$\c P5^{\cA}$ and $\c P6^{\cA}$ are the parity nodes. For each row $j$, colored symbols belong to $\mathcal D_{\c Rj}$.}
		\label{fig2}
		\vspace{-3ex}
\end{figure}

When a failure of a data node occurs, Class $\cA$ parity nodes are used to repair $\tau+1$ of the $k$ failed symbols. Class $\cA$ parity symbols are constructed in such a way that, when node $j$ is erased, $\tau+1$ data symbols in this node can be repaired reading the (non-failed) $k-1$ data symbols in the $j$-th row of the data array and $\tau+1$
parity symbols in the $j$-th row of Class $\cA$ parity nodes (see also \cref{sec:Decoding}). For later use, we define the set \c Rj as follows.
\begin{definition}
For $j=0,\ldots,k-1$, the index set $\c Rj$ is defined  as 
\[ \c Rj=\{(j,(j+1)_k),(j,(j+2)_k), \ldots,(j,(j+k-1)_k)\}. \]
\end{definition}

Then, the set $\mathcal D_{\c Rj}$ is the set of $k-1$ data symbols that are read from row $j$ to recover $\tau+1$ data symbols of node $j$ using Class $\cA$ parity nodes.
\begin{example}
An example of a Class $\cA$ code is shown in Fig.~\ref{fig2}. One can verify that the code can correct any $2$ node failures. For each row $j$, the set $\mathcal D_{\c Rj}$ is indicated in red color.  For instance, $\mathcal D_{\c R0}=\{d_{0,1},d_{0,2},d_{0,3},d_{0,4}\}$.
\end{example}

The main purpose of Class $\cA$ parity nodes is to provide good erasure correcting capability. However, the use of piggybacks helps also in reducing the number of symbols that need to be read to repair the $\tau+1$ symbols of a failed node that are repaired using the Class $\cA$ code, as compared to MDS codes. The remaining $k-\tau-1$ data symbols of the failed node can also be recovered from Class $\cA$ parity nodes, but at a high symbol read cost of $k$. 
%It should be noted that the Class $\cA$ code has a reduced repair bandwidth. It can be seen as follows. To repair $\tau+1$ symbols in a failed node, one needs to read $\tau+k$ symbols, while the remaining $k-\tau-1$ are repaired 
Hence, the idea is to add another class of parity nodes, namely Class $\cB$ parity nodes, in such a way that these symbols can be recovered with lower read cost.

\section{Class $\cB$ Parity Nodes}
\label{Sec:ClassB}

Class $\cB$ parity nodes are obtained using an $(\nB,k)$ linear block code with $\nB < 2k-\tau$ over $\text{GF}(q)$ to encode the $k\times k$ data symbols of the data array. This generates $k(\nB-k)$ Class $\cB$ parity symbols, $p^{\cB}_{i,l}$, $i=0,\ldots,k-1$, $l=\nA,\ldots,n-1$. In \cite{kum15a}, we presented an algorithm to construct Class $\cB$ codes. In this section, we present a similar construction in a much more compact, mathematical manner. 

\subsection{Definitions and Preliminaries}
\begin{definition}
\label{def:qj}
For $j=0,\ldots,k-1$, the %vector \b Qj and a respective 
	index set \c Qj is defined as 
	\begin{align*}
		&{\c Qj} = \\
		& \{((j+\tau+1)_k, j),((j+\tau+2)_k, j), \ldots, ((j+k-1)_k,j)\}.
	\end{align*}
\end{definition}
% done TODO Kill vectors and positions!!

Assume that data node $j$ fails. It is easy to see that the set $\mathcal D_{\c Qj}$ is the set of $k-\tau-1$ data symbols that are not recovered using Class $\cA$ parity nodes.
\begin{example}
\label{ex:Q0R0}
For the example in Fig.~\ref{fig2}, the set $\mathcal D_{\c Qj}$ is indicated by hatched symbols for each column $j$, $j=0,\ldots, k-1$. For instance, $\mathcal D_{\c Q0}=\{d_{2,0},d_{3,0},d_{4,0}\}$. 
\end{example}

For later use, we also define the following set.
\begin{definition}
\label{def:xj}
For $j=0,\ldots,k-1$, the index set \c Xj is defined as
\begin{align*}
\c Xj&=\{(j,(j+1)_k),(j,(j+2)_k), \ldots, (j,(j+k-\tau-1)_k)\}.
\end{align*}
\end{definition}
Note that $\c Xj=\c Rj\cap\{\cup_{l=0}^{k-1} \c Ql\}$.

\begin{example}
\label{ex:X0}
For the example in Fig.~\ref{fig2}, the set $\mathcal D_{\c Xj}$ is indicated by hatched symbols for each row $j$. For instance, $\c X0=\c R0\cap \{ \c Q0\cup \c Q1\cup \c Q2\cup \c Q3\cup \c Q4\}=\{(0,1),(0,2),(0,3)\}$, thus we have $\mathcal D_{\mathcal X_0}=\{d_{0,1},d_{0,2},d_{0,3}\}$.	
\end{example}

The purpose of Class $\cB$ parity nodes is to allow the recovery of the data symbols in $\mathcal D_{\c Qj}$, $j=0,\ldots,k-1$, at a low additional read cost. Note that after recovering $\tau+1$ symbols using Class $\cA$ parity nodes, the data symbols in the sets $\mathcal D_{\c Rj}$  are already stored in the decoder memory. Therefore, they are accessible for the recovery of the remaining $k-\tau-1$ data symbols using Class $\cB$ parity nodes without the need of reading them again. The main idea is based on the following proposition.
\begin{proposition}
\label{prep:MainIdea}
If a Class $\cB$ parity symbol $p^{\cB}$ is the sum of one data symbol $d\in \mathcal D_{\c Qj}$ and a number of data symbols in $\mathcal D_{\c Xj}$, then the recovery of $d$ comes at the cost of one additional read (one should read parity symbol $p^{\cB}$).
\end{proposition}

This observation is used in the construction of Class $\cB$ parity nodes in \cref{Sec: Construction} below to reduce the normalized repair bandwidth $\lambda$. In particular, we add up to  $k-\tau-1$ Class $\cB$ parity nodes which allow to reduce the additional read cost of all $k(k-\tau-1)$ data symbols in all $\mathcal D_{\c Qj}$'s to $1$. (The addition of a single Class $\cB$ parity node allows to recover one new data symbol in each $\mathcal D_{\c Qj}$, $j=0,\ldots,k-1$, at the cost of one additional read.)

%Consider a file that needs to be stored on a DSS. Such a file is formatted into a data 
%matrix $\bm D=[d_{i,j}]\in \text{GF}(q)^{k\times k}\, i,j=1,\ldots,k$ where each row $\bm d_i=(d_{i,0},d_{i,1},\ldots,d_{i,k-1})$ of $\bm D$ is referred to as a stripe.
%
%
% vector $\bm d=(\bm d_0,\bm d_1,\ldots, \bm d_{k-1})\in\text{GF}(q^p)^{k^2}$, where the vector $\bm d_i=(d_{0,i},d_{1,i},\ldots,d_{k-1,i})\in\text{GF}(q^p)^{k}, i=1,\ldots,k-1$ contains $k$ data symbols. In order to achieve tolerance against node failures, $\bm d$ is encoded using an $(nk,k^2)$ array erasure correcting code to obtain the codeword $\bm c=(\bm c_0,\bm c_1,\ldots, \bm c_{n-1})$, where the vector $\bm c_i=(c_{0,i},c_{1,i},\ldots,c_{k-1,i}$ consists of $k$ symbols. Each of this vector $\bm c_i$ is then stored on the respective $i^{\text{th}}$ node of the DSS.

%For two integers $a$ and $b$, $a\leq b$, we define $[a]\triangleq \{0, 1, ..., a-1\}$ and $[a,b]\triangleq\{a, a+1, a+2, ..., b-1\}$. 

 %We define a vector $\mathbf Q$ then the statement $\mathbf Q(a)=b$ means that the element b is in position a in $\mathbf Q$. If $a=\mathbf{null}$ then $\b Q{}(a)=0$.

\subsection{Construction of Class $\cB$ Nodes}
\label{Sec: Construction}
For $t=0,\ldots,k-1$, each parity symbol in the $l$-th Class $\cB$ parity node, $l=\nA,\ldots,n-1$, is sequentially constructed as 
\begin{align}
\label{Eq: ClassB_parity_const}
	p^\cB_{t,l}=d_{(\tau+1-\nA+l+t)_k,t}+\sum_{j=0}^{k-\tau-3+\nA-l} d_{t,(1+j+t)_k}.
\end{align}
The construction above follows \cref{prep:MainIdea} wherein  $d_{(\tau+1-\nA+l+t)_k,t}\in\mathcal D_{\mathcal Q_t}$ and  $\{d_{t,(1+j+t)_k}\}_{j=0}^{k-\tau-3+\nA-l}$ $\subset \mathcal D_{\mathcal X_t}$. This ensures that the read cost of each of the $k$ symbols $d_{(\tau+1-\nA+l+t)_k,t}$ is $1$. Thus, the addition of each parity node leads to $k$ data symbols to have a read cost of $1$.
Note that adding the second term in \cref{Eq: ClassB_parity_const} ensures that $k(k-\tau-1)$ data symbols are repaired by the Class $\cB$ parity nodes.
 The same principle was used in \cite{kum15a}. It should be noted that the set of data symbols used in the construction of the parity symbols in \cref{Eq: ClassB_parity_const} may be different compared to the construction in \cite{kum15a}. However,  the overall average repair bandwidth remains the same. 

\begin{remark}
For the particular case $\nB-k=k-\tau-1$ one may neglect the second term in \eqref{Eq: ClassB_parity_const}. The resulting codes would still have the same repair bandwidth and lower repair complexity than the codes built from \eqref{Eq: ClassB_parity_const}.  However, this construction would not allow  rate-compatible Class $\cB$ codes. 
%	For the particular case where $\nB-k=k-\tau-1$, we can have each parity symbol of the Class $\cB$ code to be a single unique data symbol	in $\cup_{j=0}^{k-1}\mathcal D_{\mathcal Q_j}$ by removing the second term in \cref{Eq: ClassB_parity_const}. The parity nodes thus constructed when combined with Class $\cA$ codes, achieves the same repair bandwidth but with reduced repair complexity. This is because, the parity symbols in Class $\cB$  nodes are trivial. However, such a construction makes the Class $\cB$ codes rate incompatible while the construction in \cref{Eq: ClassB_parity_const} makes the Class $\cB$ codes rate compatible.
\end{remark}

 In the sequel, we will refer to the construction of Class $\cB$ parity nodes according to \eqref{Eq: ClassB_parity_const} as Construction~$1$.

\begin{figure}[t]
	\centering
	\includegraphics{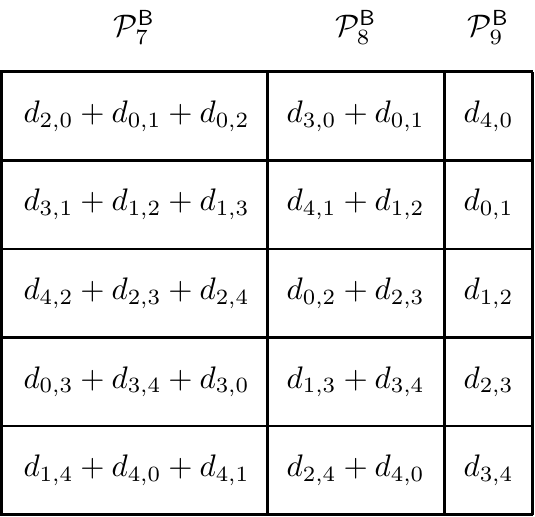}
	\caption{Class $\cB$ parity nodes for the data nodes in Fig.~\ref{fig2}.}
	\vspace{-3ex}
	\label{Fig:ClassB}
\end{figure}

\begin{example}
	\label{Ex: 85exm}
	With the aim to construct a $(10,5)$ code, consider the construction of an $(8,5)$ Class $\cB$ code where the $(7,5)$ Class $\cA$ code, with $\tau=1$, is as shown in \cref{fig2}. For $t=0,\ldots,k-1$, the parity symbols in the first Class $\cB$ parity node (the $7$-th node) are
	\ifonecolumn
	\begin{align*}
		p^\cB_{t,7}  =  d_{(2+t)_5,t}+\sum_{j=0}^{1} d_{t,(1+j+t)_5}=d_{(2+t)_5,t}+ d_{t,(1+t)_5}+d_{t,(2+t)_5}.	
	\end{align*}
	\else
	\begin{align*}
		p^\cB_{t,7}   &=  d_{(2+t)_5,t}\\
		&\;\;\;\;+\sum_{j=0}^{1} d_{t,(1+j+t)_5}=d_{(2+t)_5,t}+ d_{t,(1+t)_5}+d_{t,(2+t)_5}.	
	\end{align*}
	\fi
	The constructed parity symbols are as seen in \cref{Fig:ClassB}, where the $t$-th row in node $\mathcal P^\cB_7$ contains the parity symbol $p^\cB_{t,7}$. Notice that $d_{(2+t)_5,t}\in\mathcal D_{\mathcal Q_t}$ and $\{d_{t,(1+t)_5}, d_{t,(2+t)_5}\}\subset \mathcal D_{\mathcal X_t}$. In a similar way, the parity symbols in nodes $\mathcal P^\cB_8$ and $\mathcal P^\cB_9$ are
	\begin{align*}
		p^\cB_{t,8}&=d_{(3+t)_5,t}+\sum_{j=0}^{0}d_{t,(1+j+t)_5}=d_{(3+t)_5,t}+d_{t,(1+t)_5}
	\end{align*}
	and
	\begin{align*}	
			p^\cB_{t,9}&=d_{(4+t)_5,t}+\sum_{j=0}^{-1}d_{t,(1+j+t)_5}=d_{(4+t)_5,t},
	\end{align*}
	respectively.
	
	Consider the repair of the first data node in \cref{fig2}. The symbol $d_{0,0}$ is reconstructed using $p^\cA_{0,5}$. This requires reading the symbols $d_{0,1}$, $d_{0,2}$, $d_{0,3}$, and $d_{0,4}$. Since $p^\cA_{0,6}$ is a function of all data symbols in the first row, reading $p^\cA_{0,6}+d_{1,0}$ is sufficient for the recovery of $d_{1,0}$. From \cref{Fig:ClassB}, the symbols $d_{2,0}$, $d_{3,0}$, and $d_{4,0}$ can be recovered by reading just the parities $d_{2,0}+d_{0,1}+d_{0,2}$, $d_{3,0}+d_{0,1}$, and $d_{4,0}$, respectively. Thus, reading  $5+4=9$ symbols is sufficient to recover all the symbols in the node, and the normalized repair bandwidth is $9/5=1.8$ per failed symbol. A more formal repair procedure is presented in \cref{sec:Decoding}.

\end{example}

Adding $\nB-k$ Class $\cB$ parity nodes allows to reduce the additional read cost of $\nB-k$ data symbols from each $\mathcal D_{\c Qj}$, $j=0,\ldots,k-1$, to $1$. However, this comes at the cost of a reduction in the code rate, i.e., the storage overhead is increased. In the above example, adding $\nB-k=3$ Class $\cB$ parity nodes leads to the reduction in code rate from $R=5/7$ to $R=5/10=1/2$. If a lower storage overhead is required, Class $\cB$ parity nodes can be \textit{punctured}, starting from the last parity node (for the code in Example~\ref{Ex: 85exm}, nodes $\c P9^{\cB}$, $\c P8^{\cB}$, and $\c P7^{\cB}$ can be punctured in this order), at the expense of an increased repair bandwidth. If all Class $\cB$ parity nodes are punctured, only Class $\cA$ parity nodes would remain, and the repair bandwidth is equal to the one of the Class $\cA$ code. 
Thus, our code construction gives a family of rate-compatible codes which provides a tradeoff between repair bandwidth and storage overhead: adding more Class $\cB$ parity nodes reduces the repair bandwidth, but also increases the storage overhead.

\subsection{Repair of a Single Data Node Failure: Decoding Schedule}
\label{sec:Decoding}

The repair of a failed data node proceeds as follows. First, $\tau+1$ symbols are repaired using Class $\cA$ parity nodes. Then, the remaining symbols are repaired using Class $\cB$ parity nodes. With a slight abuse of language, we will refer to the repair of symbols using Class $\cA$ and Class $\cB$ parity nodes as the decoding of Class $\cA$ and Class $\cB$ codes, respectively. 

We will need the following definition.
\begin{definition}
\label{def:localset}
Consider a Class $\cB$ parity node and let ${\mathcal P}^{\cB}$ denote the set of parity symbols in this node. Also, let $d\in \mathcal D_{\c Qj}$ for some $j$ and $p^{\cB}\in{\mathcal P}^{\cB}$ be the parity symbol $p^{\cB}=d+\sum_{d'\in \mathcal D'} d'$, where $\mathcal D'\subset\mathcal D$, i.e., the parity symbol  $p^{\cB}$ is the sum of $d$ and a subset of other data symbols. We define $\breve{\mathcal D}=\mathcal D' \cup \{d\}$.
\end{definition}

Suppose that node $j$ fails. Decoding is as follows.
\begin{itemize}
\item \textbf{Decoding the Class $\cA$ code}. To reconstruct the failed data symbol in the $j$-th row of the code array, $k$ symbols ($k-1$ data symbols and $p^{\cA}_{j,k}$) in the $j$-th row are read. These symbols are now cached in the memory. We then read  the $\tau$ piggybacked symbols in the $j$-th row. By construction (see (\ref{eq:piu})), this allows to repair $\tau$ failed symbols, at the cost of an additional read each.
\item \textbf{Decoding the Class $\cB$ code}. Each remaining failed data symbol $d_{i,j}\in\mathcal D_{\mathcal Q_j}$ is obtained by reading a Class $\cB$ parity symbol whose corresponding set $\breve{\mathcal D}$ (see Definition~\ref{def:localset}) contains $d_{i,j}$. In particular, if several Class $\cB$ parity symbols $p^{\cB}_{i',j'}$ contain $d_{i,j}$, we read the parity symbol with largest index $j'$. This yields the lowest additional read cost.
\end{itemize}

\section{A Heuristic Construction of Class $\cB$ Nodes With Improved Repair Bandwidth} 
\label{sec:improvement}

In this section, we provide a way to  improve the repair bandwidth of the family of codes constructed so far. More specifically, we achieve this by  providing a heuristic algorithm for the construction of the Class $\cB$ code, which improves Construction~$1$ in \cref{Sec:ClassB} for some values of $n$ and  even values of $k$.

The algorithm is based on a simple observation. Let $p^\cB_1$ and $p^\cB_2$ be two parity symbols constructed from $\rho$ data symbols in $\mathcal D$ in two different ways as follows:
\begin{align}
\label{Eq: Opt_Obs}
	p^\cB_1&=d_{i,j}+d_{j,i}+d_{j,i_2}+\cdots+d_{j,i_{\rho-1}},\\
\label{Eq: Opt_Obs2}
	p^\cB_2&=d_{i,j}+d_{j,i_1}+d_{j,i_2}+\cdots+d_{j,i_{\rho-1}},
\end{align}
where $d_{i,j}\in\mathcal D_{\mathcal Q_j}$ (see \cref{def:qj}), $i_1,\ldots, i_{\rho-1}\not=i$, and $d_{j,i_1},d_{j,i_2},\ldots,d_{j,i_{\rho-1}}\in\mathcal D_{\mathcal X_j}$ (see \cref{def:xj}). Note that the only difference between the two parity symbols above is that $p^\cB_2$ does not involve $d_{j,i}$ (and that  $p^\cB_1$ does not involve $d_{j,i_1}$).  This has a major consequence in the repair of the data symbols $d_{i,j},d_{j,i},\ldots,d_{j,i_\rho-1}$ and $d_{i,j},d_{j,i_1},\ldots,d_{j,i_\rho-1}$ using $p^\cB_1$ and $p^\cB_2$, respectively. Consider the repair using parity symbol $p^\cB_1$. From \cref{prep:MainIdea}, it is clear that the repair of symbol $d_{i,j}$ will have an additional read cost of $1$, since the remaining $\rho-1$ data symbols are in $\mathcal D_{\mathcal X_j}$. As the symbol $d_{j,i}\in\mathcal D_{\mathcal Q_i}$ and $d_{i,j}\in \mathcal D_{\mathcal X_i}$, from Proposition~\ref{prep:MainIdea} and the fact that $d_{j,i_2},\ldots,d_{j,i_{\rho-1}}\not\in\mathcal D_{\mathcal X_i}$, we can repair $d_{j,i}$ with an additional read cost of $\rho-1$. The remaining $\rho-2$ symbols each have an additional read cost of $\rho$, whereas the symbols repaired using $p^\cB_2$ incur an additional read cost of $1$ for the symbol $d_{i,j}$ and $\rho$ for the remaining symbols. Clearly, we see that the combined additional read cost, i.e., the sum of the individual additional read costs for each data symbol using $p^\cB_1$ is lower (by $1$) than that using $p^\cB_2$.
%When $d_{j,i}\not\in\mathcal D_{\mathcal X_j}$, clearly, the difference between the two ways of constructing the parity symbol is that in the former of the $\rho$ data symbols, two symbols have read cost of $1$ and $\rho-1$ and the remaining $\rho-2$ symbols having a cost of $\rho$. While the latter construction has $1$ data symbol with a read cost of $1$ and the remaining $\rho-1$ symbols having a cost of $\rho$. Therefore, the former construction leads to reduction in read cost by $1$. 

In the way Class $\cA$ parity nodes are constructed and due to the structure of the sets $\mathcal D_{\mathcal Q_j}$ and $\mathcal D_{\mathcal X_j}$, it can be seen that $d_{i,j}\in\mathcal D_{\mathcal Q_j}$ and $d_{j,i}\in\mathcal D_{\mathcal X_j}$ when $k\ge 2(\tau+1)$. From Construction~$1$ of the Class $\cB$ code in \cref{Sec:ClassB} we observe that for odd $k$ and  $k>2(\tau+1)$, the parity symbols in node $\mathcal P^\cB_l$ are as in \cref{Eq: Opt_Obs} for $\nA\leq l\leq \nA+\lfloor k/2\rfloor-\tau-1$. Furthermore, for $\nA+\lfloor k/2\rfloor-\tau\le l \le n-1$,  the parity symbols in node $\mathcal P^\cB_{l}$ have the structure in \cref{Eq: Opt_Obs2}.  On the other hand, for $k$ even and $k\geq2(\tau+1)$, the parity symbols in the node $\mathcal P^\cB_l$ are as in \cref{Eq: Opt_Obs} for  $\nA\leq l\leq \nA+k/2-\tau-2$. However, contrary to case of $k$ odd, the parity symbols in the node $\mathcal P^\cB_{\nA+k/2-\tau-1}$ follow \cref{Eq: Opt_Obs2}. But since $k\geq2(\tau+1)$, we know that $d_{i,j}\in\mathcal D_{\mathcal Q_j}$ and $d_{j,i}\in\mathcal D_{\mathcal X_j}$. Thus, it is possible to construct some parity symbols in this node  as in \cref{Eq: Opt_Obs}, and Construction~$1$ of Class $\cB$ nodes in the previous section can be improved. However, the improvement comes at the expense of the loss of the mathematical structure of Class $\cB$ nodes given in \eqref{Eq: ClassB_parity_const}. 

\begin{figure}
	\centering
\includegraphics{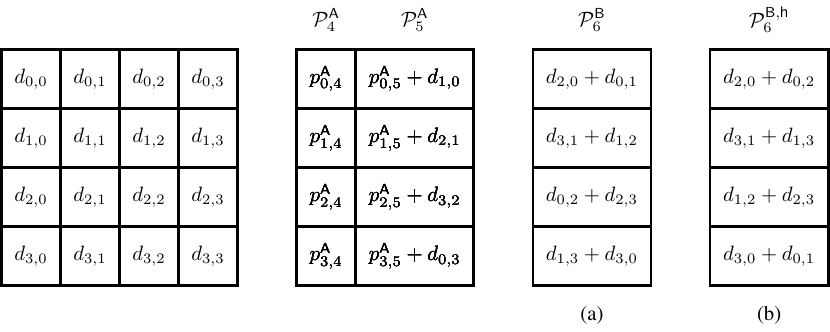}
\caption{A $(7,4)$ code constructed from a $(6,4)$ Class $\cA$ code with $\tau=1$ and a $(5,4)$ Class $\cB$ code. (a) Class $\cB$ node constructed according to Construction~$1$ in \cref{Sec:ClassB}. (b) A different configuration of the Class $\cB$ node that reduces the repair bandwidth.}
\label{Fig: ex_Code_opt}
\end{figure}

\begin{example}
	Consider the $(7,4)$ code as shown in \cref{Fig: ex_Code_opt}. \cref{Fig: ex_Code_opt}(a) shows the node $\mathcal P^\cB_6$ using Construction~$1$ in \cref{Sec:ClassB}, while \cref{Fig: ex_Code_opt}(b) shows a different configuration of the node $\mathcal P^\cB_6$. Note that $k=2(\tau+1)=4$. Thus, each pair $(\mathcal D_{\mathcal Q_j},\mathcal D_{\mathcal X_j})$ contains one symbol $d_{i,j}$ and $d_{j,i}$. The node $\mathcal P^\cB_6$ has parity symbols according to \cref{Eq: Opt_Obs2}, while $\mathcal P^{\cB, \mathsf{h}}_6$ has two parity symbols as in \cref{Eq: Opt_Obs} and two parity symbols according to \cref{Eq: Opt_Obs2}. The configuration of the $(7,4)$ code arising from Construction~$1$ has a normalized repair bandwidth of $2$, while the $(7,4)$ code with node $\mathcal P^{\cB,\mathsf{h}}_6$ in \cref{Fig: ex_Code_opt}(b)  has a repair bandwidth of $1.875$, i.e., an improvement is achieved.
\end{example}

% In fact, when $k>2(\tau+1)$ construction of Class $\cB$ parity symbols in \cref{Sec:ClassB} follows \cref{Eq: Opt_Obs}. It is when $k=2(\tau+1)$ that the parity symbols no longer follow \cref{Eq: Opt_Obs}. In other words, $k$ is even. 
% An example of such a case is seen in \cref{Fig: ex_Code_opt}. \cref{Fig: ex_Code_opt}(a) shows the original construction that has a repair bandwidth of $2$ whereas the Class $\cB$ node in \cref{Fig: ex_Code_opt}(b) $\mathcal P^{\cB,\text{red}}_6$ induces a reduced repair bandwidth of $1.825$. Notice that the node, $\mathcal P^{\cB,\text{red}}_6$ in has two parity symbols that constructed using the template in \cref{Eq: Opt_Obs} while none of the parity symbols in the node $\mathcal P^{\cB}_6$ follow it.

In order to describe the modified code construction, we define the function $\mathrm{read}(d, p^\cB)$ %and $\text{position}(a, \mathbf Q)$ 
as follows. 
\begin{definition}
\label{def:read}
%Consider a Class $\cB$ parity node and denote by ${\mathcal P}^{\cB}$ the set of parity symbols in this node. Let $p\in{\mathcal P}^{\cB}$ be $p=d+\sum_{i=1}^{m} d'_{i}$, where $d\in \mathcal D_{\c Qj}$ for some $j$ and $d_i'\in \mathcal D'\subset\mathcal D$. Then,
%Consider a Class $\cB$ parity node and let ${\mathcal P}^{\cB}$ denote the set of parity symbols in this node. Also, let $d\in \mathcal D_{\c Qj}$ for some $j$ and $p^{\cB}\in{\mathcal P}^{\cB}$ be the parity symbol $p^{\cB}=d+\sum_{d'\in \mathcal D'} d'$, where $\mathcal D'\subset\mathcal D$. Then,
Consider the construction of the parity symbol $p^{\cB}$ as $p^{\cB}=d+\sum_{d'\in \mathcal D'} d'$ (see \cref{def:localset}). Then,
	\begin{align*}
		\readd(d,p^{\cB})=| \breve{\mathcal D}\backslash \mathcal D_{\c Xj} |.
	\end{align*}
%where $\breve{\mathcal D}$ is defined in .
\end{definition} 
%TODO Add a remark for the case when d is not in p^b.
%\begin{definition}
%	For a vector $\mathbf Q$, we define $position(d,\mathcal Q)$ as the position of block $d$ in that vector. If $d$ is not in $\mathbf Q$, $position=\textbf{null}$.
%\end{definition}

For a given data symbol $d$, the function $\readd(d,p^{\cB})$ gives the additional number of symbols that need to be read to recover $d$ (considering the fact that some symbols are already cached in the memory). The set $\breve{\mathcal D}$ represents the set of data symbols that the parity symbol $p^\cB$ is a function of. We use the index set $\mathcal U$ to represent the indices of such data symbols. We denote by $\mathcal U_t,t=0,\ldots,k-1$, the index set corresponding to the $t$-th parity symbol in the node (there are $k$ parity symbols in a parity node). 

In the following, denote by $\A = [a_{i,j}]$ a temporary matrix of read costs for the respective data symbols in $\D = [d_{i,j}]$. After Class $\cA$ decoding,
\begin{align}
	a_{i,j}=
	\begin{cases}
		\infty & \text{if } d_{i,j}\in \cup_{t=0}^{k-1} \mathcal D_{\mathcal{Q}_t}\\
		k & \text{if } i=j\\
		1 &  \text{otherwise}
	\end{cases}.
	\label{Eq:A}
\end{align}
%where $t=0,\ldots,k-1$. 
In Section~\ref{Sec:ConstEx} below, we will show that the construction of parities depends upon the entries of $\bm A$. To this extent, for some real matrix $\bm M = [m_{i,j}]$ and index set $\mathcal I$, we define $\Psi(\bm M_{\mathcal I})$ as the set of indices of matrix elements of $\bm M$ from $\mathcal I$ whose values are equal to the maximum of all entries in $\bm M$ indexed by $\mathcal I$. More formally, $\Psi(\bm M_{\mathcal I})$ is defined as
\begin{align*}
	\Psi(\bm M_{\mathcal I})=\left\{(i,j)\in\mathcal I\mid m_{i,j}=\max_{(i',j')\in\mathcal I} m_{i',j'} \right\}.
\end{align*}

The heuristic algorithm to construct the Class $\cB$ code is given in Appendix~\ref{App: Algorithm} and we will refer to the construction of the Class $\cB$ code according to this algorithm as Construction~$2$. In the following subsection, we clarify the heuristic algorithm to construct the Class $\cB$ code with the help of a simple example. 

%
%defined as
%\begin{align} 
%	\Psi(\bm M)=\{(i',j')|(i',j')=\argmax_{\forall i,j}m_{i,j},\bm M=[m_{i,j}]\},
%\end{align}
%where $\bm M=[m_{i,j}]$ is an arbitrary matrix. The function $\Psi(\bm M)$ outputs an index set that corresponds to maximum values of the elements in $\bm M$.

%Since $a_{i,j}$ represents the read cost for the data symbol $d_{i,j}$, there is a one-to-one correspondence between the two. We define a bijection between the set of data symbols $\mathcal D$ and $\mathcal A$, i.e., the set of elements of $\bm A$ as follows
%\begin{equation} \label{Eq: Mapping}
%\begin{split} 
%%\begin{align}
%	f: \mathcal D &\rightarrow\mathcal A  \\
%	   d_{i,j}  &\mapsto a_{i,j}. 
%\end{split}
%\end{equation}

%\vspace{-1cm}

\subsection{Construction Example}
\label{Sec:ConstEx}

Let us consider the construction of a $(7,4)$ code using a $(6,4)$ Class $\cA$ code and a $(5,4)$ Class $\cB$ code. In total, there are three parity nodes; two Class $\cA$ parity nodes, denoted by $\mathcal P^\cA_4$ and $\mathcal P^\cA_5$, respectively, and one Class $\cB$ parity node, denoted by  $\mathcal P^{\cB,\mathsf{h}}_6$, where the upper index $\mathsf{h}$ is used to denote that the parity node is constructed using the heuristic algorithm.  The parity symbols of the nodes are depicted in \cref{Fig: ex_Code_opt}. Each parity symbol of the  Class $\cB$ parity node is the sum of $k-\tau-1$ data symbols $d_{i,j}\in\cup_{j'}\mathcal D_{\mathcal Q_{j'}}$, constructed such that the read cost of each symbol $d_{i,j}$ is lower than $a_{i,j}$ as shown below.

%The parity nodes $\mathcal P^\cA_k,\ldots,\mathcal P^\cA_{\nA-1},\mathcal P^\cB_l,l=\nA,\ldots,\nA+k/2-\tau-2$ comes from the Class $\cA$ code and Class $\cB$ codes as in \cref{Sec:ClassA,Sec:ClassB}. After the construction of the parity nodes $\bm A$ is updated with the read values for the symbols $d_{i,j}$ by using the parity symbols $p_{i,l}\in\mathcal P^\cB_l$. Then, the algorithm uses $\bm A$ and recursively construct parity symbols for the remaining parity nodes such that the read costs of symbols $d_{i,j}$ are the lowest possible. Further, each subsequent parity node has parity symbols that are  constructed with one less data symbol than the parity symbols in the last node.

%The remaining nodes are constructed  recursively as follows:
\begin{itemize}
	\item[\textbf{1.}] \textbf{Construction of} $\mathcal P^{\cB,\mathsf{h}}_6$\\
	Each parity symbol in this node is constructed using $\rho_6=k-\tau-1=2$ unique symbols as follows.
	\begin{enumerate}
		\item[1.a] Since no symbols have been constructed yet, we have $\mathcal U_{t_1}=\emptyset$, $t_1=0,\ldots,3$. (This corresponds to the execution of \cref{alg:SeedConstruction_P1s} to \cref{alg:SeedConstruction_P1e} of \cref{alg:SeedConstruction} in Appendix~\ref{App: Algorithm}.)
		\item[1.b] Select $d_{i,0}\in\mathcal{D}_{\mathcal Q_0}$ such that its read cost is maximum, i.e., $d_{i,0}\in\mathcal D_{\Psi(\bm A_{\mathcal Q_0})}$. Choose $d_{i,0}=d_{2,0}$, as $a_{2,0}=\infty$. Note that we choose $d_{2,0}$ since $d_{0,2}\in\mathcal D_{\mathcal X_0}$.
		\item[1.c] Construct  $p_{0,6}=d_{2,0}+d_{0,2}$ (see \cref{alg:SeedConstruction_Choose_ji} of Algorithm~\ref{alg:SeedConstruction}). Correspondingly, we have $\mathcal U_0=\{(2,0),(0,2)\}$.
		\item[1.d] Recursively construct the next parity symbol in the node as follows. Similar to Item 1.b, choose $d_{3,1}\in\mathcal D_{\Psi(\bm A_{\mathcal Q_1})}$.  Construct $p_{1,6}=d_{3,1}+d_{1,3}$. Likewise, we have $\mathcal U_1=\{(3,1), (1,3)\}$
		\item[1.e] For the next parity symbol, note that $d_{0,2}$ is already used in the construction of $p_{0,6}$. The only possible choice of symbol in $\mathcal D_{\mathcal Q_2}$ is $d_{1,2}$, but $d_{2,1}\not\in\mathcal D_{\mathcal X_2}$. Therefore, we choose $d_{i_1,2}\in\mathcal D_{\Psi(\bm A_{\mathcal Q_2 \setminus \cup_{j'} \mathcal{U}_{j'}})}$ (see Line~7 of Algorithm~\ref{alg:SeedConstruction}).
In particular, since $a_{1,2}=\infty$, we choose $d_{i_1,2}=d_{1,2}$. Then, Lines~8 to 11 of Algorithm~\ref{alg:SeedConstruction} are executed.		
		\item[1.f] Choose an element $d_{2,i_2}\in\mathcal D_{\mathcal X_2\backslash\cup_{j'}\mathcal U_{j'}}$. In other words choose a symbol in $\mathcal D_{\mathcal X_2}$ which has not been used in $p_{0,6}$ and $p_{1,6}$. We have $d_{2,i_2}=d_{2,3}$. Construct $p_{2,6}=d_{1,2}+d_{2,3}$. Thus, $\mathcal U_2=\{(1,2),(2,3)\}$.
		\item[1.g] To construct the last parity symbol, we look for data symbols from the sets $\mathcal D_{\mathcal Q_3}$ and $\mathcal D_{\mathcal X_3}$. However, all symbols in $\mathcal D_{\mathcal Q_3}$ have been used in the construction of previous parity symbols. Therefore, we cyclically shift to the next pair of sets $(\mathcal D_{\mathcal Q_0},\mathcal D_{\mathcal X_0})$. Following Items 1.e and 1.f, we have $p_{3,6}=d_{3,0}+d_{0,1}$ and $\mathcal U_3=\{(3,0),(0,1)\}$.
	\end{enumerate}
%	\item[\textbf{2.}] \textbf{Construction of } $\mathcal P^{\cB,\text{red}}_7$\\
%	The parity symbols in this node is constructed using $k-\tau-2=1$ symbol.  Each parity symbol is in itself the replicated version of the data symbol that had the  
	
%	\begin{align}
%		p^\cB_{j,7}=d_{i,j}\in\mathcal D_{\mathcal Q_j}, j=0,\ldots,k-1,
%	\end{align}
%	where the symbol $d_{i,j}$ is 
%	\begin{align}
%		\argmax_{d_{i,j}\in\cup_j \mathcal D_{\mathcal Q_j}\backslash(\cup_{j_1}\hat{\mathcal D}_{j_1,7})} \text{read}(d_{i,j},\mathcal P^{\cB,\text{red}}_6), 
%	\end{align}
%	and the corresponding $a_{i,j}>1$.
\end{itemize}
Note that $|\mathcal U_t|=2$ for all $t$, thus this completes the construction of the $(7,4)$ code. The Class $\cB$ parity node constructed above is depicted in \cref{Fig: ex_Code_opt}(b). 

\subsection{Discussion}
%\begin{remark}
	In general, the algorithm constructs $\nB-k$ parity nodes, $\mathcal P^\cB_{\nA},\ldots,\mathcal P^\cB_{n-1}$, recursively. In the $l$-th Class $\cB$ node, $l=\nA,\ldots,n-1$, each parity symbol is a  sum of at most $\rho_l=k-\tau-1-l+\nA$ symbols $d_{i,j}\in\cup_{j'}\mathcal D_{\mathcal Q_{j'}}$. Each parity symbol $p_{t,l}$, $t=0,\ldots,k-1$, in the $l$-th Class $\cB$ parity node with $\rho_l>1$ is constructed recursively with $\rho_l-1$ recursion steps. In the first recursion step, each parity symbol $p_{t,l}$ is either equal to a single data symbol or a sum of $2$ data symbols. In the latter case, the first symbol $d_{i,j}\in\mathcal D_{\mathcal Q_t}$  is chosen as the symbol with the largest read cost $a_{i,j}$. The second symbol is $d_{j,i}\in\mathcal D_{\mathcal X_t}$ if such a symbol exists. Otherwise (i.e., if $d_{j,i}\not\in\mathcal D_{\mathcal X_t}$), symbol $d_{j,i''}\in\mathcal D_{\mathcal X_t}$ is chosen. In the remaining $\rho_l-2$ recursion steps a subsequent data symbol $d_{j,i'}\in\mathcal D_{\mathcal X_t}$  (if it exists) is added to $p_{t,l}$. Doing so ensures that $k$ symbols have a new read cost that is reduced to $1$ when parity symbols $p_{t,l}$ are used to recover them. Having obtained these parity symbols, the read costs of all data symbols in $\cup_{j'}\mathcal D_{\mathcal Q_{j'}}$ are updated and stored in $\bm A$. This process is repeated for successive parity nodes. If $\rho_l=1$ for the  $l$-th parity node, its parity symbols $p_{t,l}$ are equal to the data symbols $d_{i,j}\in\mathcal D_{\mathcal Q_t}$ whose read costs $a_{i,j}$ are the maximum possible.
	
	%In the above example, since $\theta=2$, we only see a single recursion for the construction of $\mathcal P^{\cB,\text{red}}_6$, where each parity symbol is a sum of two data symbols.
	In the above example, only a single recursion for the construction of $\mathcal P^{\cB,\mathsf{h}}_6$ is needed, where each parity symbol is a sum of two data symbols. 
%\end{remark}

\section{Code Characteristics and Comparison} \label{Sec:Comparison}
%\begin{adjustbox}{width=1.2\textwidth,center}
\begin{table*}[h]
%\scriptsize
	\centering
	\caption{\footnotesize Comparison of $(n,k)$ codes that aim at reducing the repair bandwidth. The repair bandwidth and the repair complexity are normalized per symbol, while the encoding complexity is given per row of the code array.  Note that for MDR and EVENODD codes, $n=k+2$ where $k$ is prime for EVENODD codes.}
	\label{Tab:Summary}
	\def\Hline{\noalign{\hrule height 2\arrayrulewidth}}
    \vskip -2.0ex %take space out between caption and table
	\begin{tabular}{@{}p{1.9 cm}cp{2.0 cm}ccc@{}}
	\Hline	
			& $\beta$ & \textbf{Fault Tolerance} & \textbf{Norm. Repair Band.} & \textbf{Norm. Repair Compl.} & \textbf{Enc. Complexity}\\
	\hline \\ [-2.0ex] \hline  \\ [-2.0ex]
			MDS & $1$ & \centering $n-k$ & $k$ & $O((k-1)\nu+k\nu^2)$ & $O((n-k)(k-1)\nu+k\nu^2)$\\[5 pt]
			LRC \cite{hua12} & $1$ & \centering $r+1$ & $\frac{k}{n-k-r}$ & $O((\lceil\frac{k}{n-k-r}\rceil-1)\nu)$ & \scalebox{0.72}{$rO((k-1)\nu+k\nu^2)+(n-k-r)O((\lceil\frac{k}{n-k-r}\rceil-1)\nu)$}\\[5 pt]
			MDR \cite{wan14} & $2^k$ & \centering $2$ & $\frac{k+1}{2}$ & $O(k-1)$ & $O(2(k-1))$\\[5 pt]
			Zigzag \cite{tam13} & $(n-k)^{k-1}$ & \centering $n-k$ & $\frac{n-1}{n-k}$ &$O((k-1)\nu+k\nu^2)$ & $O((n-k)(k-1)\nu+k\nu^2)$\\[5 pt]
			Piggyback \cite{Ras17} & $2$ & \centering $n-k$ & \scalebox{0.9}{$\frac{(k-t_r)(k+t)+t_r(k+t_r+\ell-2)}{2k}$} & -- & --\\[5 pt]
			EVENODD\cite{Bla95} & $k-1$ & \centering$2$ & $k$ & $O((k-1)\nu)$ & $O(\frac{2k^2-2k-1}{k-1}\nu)$\\[5pt]
			Proposed codes & $k$ & \centering $f$ & $\lambda^{\mathsf s}$ & $C_{\mathsf{r}}^{\mathsf s}/k$ & $C_{\mathsf{e}}$\\[5 pt]
 		\hline
	\end{tabular}  
	\vspace{-2ex}
\end{table*}
%\end{adjustbox}

In this section, we characterize different properties of the codes presented in Sections~\ref{Sec:ClassA}-\ref{sec:improvement}. In particular, we focus on the fault tolerance, repair bandwidth, repair complexity, and encoding complexity. We consider the complexity of elementary arithmetic operations on field elements of size $\nu = m \lceil \log_2 p \rceil$ in $\text{GF}(q)$, where $q=p^m$ for some prime number $p$ and positive integer $m$. The complexity of addition is $O(\nu)$, while that of multiplication is $O(\nu^2)$, where the argument of $O(\cdot)$ denotes the number of elementary binary additions.\footnote{It should be noted that the complexity of multiplication is quite pessimistic. However, for the sake of simplicity we assume it to be $O(\nu^2)$. When the field is $\mathrm{GF}(2^\nu)$ there exist algorithms such as the Karatsuba-Ofman algorithm \cite{Kar63,Wei06} and the Fast Fourier Transform \cite{Pol71, Cra01,Gao10} that lower the complexity to $O(\nu^{\log_2 3})$ and $O(\nu \log_2 \nu)$, respectively.}

%All properties are summarized in Table \ref{Tab:Summary} and discussed below.

\subsection{Code Rate}
\label{sec:rate}
The code rate for the proposed codes is given by $R=\frac{k}{\nA+\nB-k}$. It can be seen that the code rate is bounded as
\begin{align*}
	\frac{k}{3k-\tau-2}\leq R\leq \frac{k}{k+3}.
\end{align*}
The upper bound is achieved when $\nA=k+2$, $\tau=1$, and $\nB=k+1$, while the lower bound is obtained from the upper bounds on $\nA$ and $\nB$ given in \cref{Sec:ClassA,Sec:ClassB}.

\subsection{Fault Tolerance}
\label{sec:tolerance}

The proposed codes have fault tolerance equal to that of the corresponding Class $\cA$ codes, which depends on the MDS code used in their construction and $\tau$ (see \cref{th:fault_tol}). 
Class $\cB$ nodes do not help in improving the fault tolerance. The reason is that improving the fault tolerance of the Class $\cA$ code requires the knowledge of the piggybacks that are strictly not in the set $\cup_j \mathcal{D}_{Q_j}$, while Class $\cB$ nodes can only be used to repair symbols in $\cup_j \mathcal{D}_{Q_j}$.% while the repair of a node requires repairing data symbols outside the set $\cup_j \mathcal Q_j$. %When these nodes are combined with Class $\cA$ parity nodes, the same reason prevents in improving the fault tolerance

In the case where the Class $\cB$ code has parameters $(\nB=k+1,k)$, the resulting $(n=\nA+1,k)$ code has fault tolerance $\nA-k$ for $\tau<\xi$, i.e., one less than that of an $(n=\nA+1,k)$ MDS code.

\subsection{Repair Bandwidth of Data Nodes}
%Let the codes be constructed from a $(\nA, k)$ Class $\cA$ and $(\nB, k)$ Class $\cB$ code. 

According to \cref{sec:Decoding}, to repair the first $\tau+1$ symbols in a failed node, $k-1$ data symbols and $\tau+1$ Class $\cA$ parity symbols are read. The remaining $k-\tau-1$ data symbols in the failed node are repaired by reading Class $\cB$ parity symbols. 

Let $f_l$, $l=\nA,\ldots,n-1$, denote the number of parity symbols that are used from the $l$-th Class $\cB$ node according to the decoding schedule in Section~\ref{sec:Decoding}. Due to Construction~$1$ in \cref{Sec:ClassB}, we have
\begin{align}
\label{Eq: f_l}
\begin{split}
	f_{\nA} &=f_{\nA+1}=\cdots=f_{n-2}=1,\\
	f_{n-1}&=k-\tau-1-\sum_{l=\nA}^{n-2}f_l=k-\tau-n+\nA.
\end{split}
\end{align}
The Class $\cB$ nodes $\nA,\ldots,n-2$ are used to repair $n-1-\nA$ symbols with an additional read cost of $n-1-\nA$ ($1$ per symbol). The remaining $k-\tau-n+\nA=f_{n-1}$ erased symbols are corrected using the $(n-1)$-th Class $\cB$ node. The repair of one of the $f_{n-1}$ symbols entails an additional read cost of $1$. On the other hand, since the parity symbols in the $(n-1)$-th Class $\cB$ node are a function of $f_{n-1}$ symbols, the repair of the remaining $f_{n-1}-1$ symbols entails an additional read cost of at most $f_{n-1}$ each. In all, the $k-\tau-1$ erased symbols in the failed node have a total additional read cost of at most $n-\nA+(f_{n-1}-1)f_{n-1}$. The normalized repair bandwidth for the failed systematic node is therefore given as
\begin{align*}
\scalemath{0.95}{
	\lambda^{\mathsf s}\leq\frac{k+\tau+n-\nA+(f_{n-1}-1)f_{n-1}}{k}=\frac{2k-2f_{n-1}+f_{n-1}^2}{k}.}
\end{align*}
%
%As seen in \cref{Sec:ClassB}, the parity symbols in the first Class $\cB$ parity node are constructed from sets of data symbols of cardinality $| \mathcal D_{\mathcal{Q}_{j}}|=k-\tau-1$. Therefore, to repair each of the $k-\tau-1$ data symbols in this set it is sufficient to read at most $k-\tau-1$ symbols. The remaining Class $\cB$ parity nodes are constructed from fewer symbols than $k-\tau-1$. An upper bound on the normalized repair bandwidth is therefore
%\begin{align*}
%\lambda\leq(k+\tau+(k-\tau-1)^2)/k.
%\end{align*}
%This bound is strict and meets with equality when $k$ is odd. The reason for this follows from the similar intuition in \cref{sec:improvement}. 
Note that $f_{n-1}$ is function of $\tau$, and it follows from \cref{sec:tolerance,Eq: f_l} that when $\tau$ increases, the fault tolerance reduces while $\lambda^{\mathsf s}$ improves. Furthermore, as $\nB$ increases (thereby as $n$ increases), $f_{n-1}$ decreases. This leads to a further reduction of the normalized repair bandwidth.

\subsection{Repair Complexity of a Failed Data Node}

To repair the first symbol requires $k$ multiplications and $k-1$ additions. To repair the following $\tau$ symbols require an additional $\tau k$ multiplications and additions. Thus, the repair complexity of repairing $\tau+1$ failed symbols is
\begin{align*}
	C_{\mathsf r}^{\cA}=O((k-1)\nu+k\nu^2)+O(\tau k(\nu+\nu^2)).
\end{align*}

For Construction~$1$, the remaining $k-\tau-1$ failed data symbols in the failed node are corrected using $k-\tau-1$ parity symbols from $\nB-k$ Class $\cB$ nodes. To this extent,
% let $f_l$, $l=\nA,\ldots,n-1$, denote the number of parity symbols that are used from the $l$-th Class $\cB$ node. 
 note that  $\sum_{l=\nA}^{n-1}f_l=k-\tau-1$. The repair complexity for repairing the remaining $k-\tau-1$ symbols is
\begin{align} \label{eq:CrB}
	C_{\mathsf r}^{\cB}=\sum_{l=\nA}^{n-1} O(f_l(k-\tau -2 -  l+\nA  )\nu).
\end{align} 
% According to Construction~1 in \cref{Sec:ClassB}, we have
% \begin{align*}
% \begin{split}
% 	f_{\nA} &=f_{\nA+1}=\cdots=f_{n-2}=1,\\
% 	f_{n-1}&=k-\tau-1-\sum_{l=\nA}^{n-2}f_l.
% \end{split}
% \end{align*}
 From \cref{Eq: f_l},  (\ref{eq:CrB}) simplifies to
\ifonecolumn
\begin{align*}
% \begin{split} 
	C_{\mathsf r}^{\cB} =\sum_{l=\nA}^{n-2} O((k-\tau -2 -  l+\nA  )\nu)+  O((k-\tau-n+\nA)(k-\tau -1 -  n+\nA  )\nu).
%	 \end{split} 
\end{align*}
\else
\begin{align*}
 \begin{split} 
	C_{\mathsf r}^{\cB}&=\sum_{l=\nA}^{n-2} O((k-\tau -2 -  l+\nA  )\nu) \\
	&\;\;\;\;+O((k-\tau-n+\nA)(k-\tau -1 -  n+\nA  )\nu)\\
	 	&=O\bigg(\frac{1}{2}(n-1-\nA)(2k-2\tau-2+\nA-n)\nu\bigg)\\
	 	&\;\;\;\;+O((k-\tau-n+\nA)(k-\tau -1 -  n+\nA  )\nu).
	 \end{split} 
\end{align*}
\fi 

For Construction~$2$, the final $k-\tau-1$ failed data symbols require at most $k-\tau-2$ additions, since Class $\cB$ parity symbols are constructed as sums of at most $k-\tau-1$ data symbols. The corresponding repair complexity is therefore
 \begin{align*}
C^{\mathsf{B}}_{\mathsf{r}} &\leq O((k-\tau-2)(k-\tau-1)\nu).
\end{align*}

Finally, the total repair complexity is $C_{\mathsf{r}}^{\mathsf s}=C_{\mathsf r}^{\cA}+C_{\mathsf r}^{\cB}$.

% To repair the first symbol requires $k$ multiplications and $k-1$ additions. To repair the following $\tau$ symbols require an additional $\tau k$ multiplications and additions. The final $k-\tau-1$ symbols require at most $k-\tau-2$ additions, since Class $\cB$ parity symbols are constructed as sums of at most $k-\tau-1$ data symbols. The repair complexity of one failed node is therefore
% \begin{align*}
%C_{\textrm{R}} &\leq  O((k-1)\nu+k\nu^2)+O(\tau k(\nu+\nu^2))\\
%&\;\;\;\;+O((k-\tau-2)(k-\tau-1)\nu),
%\end{align*}
%where the first two terms correspond to the Class $\cA$ code, while the last term corresponds to the Class $\cB$ code.

\subsection{Repair Bandwidth and Complexity of Parity Nodes}
We characterize the normalized repair bandwidth and repair complexity of Class $\cA$ and $\cB$ parity nodes. 

Class $\cA$ nodes consist of $\nA-k$ MDS parity nodes of which $\tau$ nodes are modified with a single piggyback. Thus, the repair of each parity symbol in the $\nA-k-\tau$ non-modified nodes requires  downloading $k$ data symbols. To obtain the parity symbol, one needs to perform $k-1$ additions and $k$ multiplications. Thus, each parity symbol in these nodes has a repair bandwidth of $k$ and a repair complexity of $O((k-1)\nu+k\nu^2)$,  while each erased parity symbol in the $\tau$ piggybacked nodes requires reading $k+1$ data symbols. Such parity symbols are obtained by performing $k-1$ additions and $k$ multiplications to get the original MDS parity symbol and then  finally a single addition of the piggyback to the MDS parity symbol is required. Overall, the normalized repair bandwidth is $k+1$ and the normalized repair complexity is  $O(k(\nu+\nu^2))$. In average, the normalized repair bandwidth and the normalized repair complexity of Class $\cA$ parity nodes are
\ifonecolumn
\begin{align*}
	\lambda^{\mathsf{p}, \cA}=k+\frac{\tau}{\nA-k},\,\,\,\, 	C^{\mathsf{p}, \cA}_{\textrm{R}}=O\bigg((k-1)\nu+k\nu^2+\frac{\tau\nu}{\nA-k}\bigg),
\end{align*}
\else
\begin{align*}
	\lambda^{\mathsf{p}, \cA}&=k+\frac{\tau}{\nA-k},\\
	C^{\mathsf{p}, \cA}_{\mathsf{r}}&=O\bigg((k-1)\nu+k\nu^2+\frac{\tau\nu}{\nA-k}\bigg),
\end{align*}
\fi
respectively.

Considering the $i$-th Class $\cB$ node, the repair of an erased parity symbol requires downloading $k-\tau-1-i$, $i=0,\ldots,\nB-k-1$, data symbols. The repair entails $k-\tau-2-i$ additions, and the average normalized repair bandwidth $\lambda^{\mathsf{p}, \cB}$ and repair complexity $C^{\mathsf{p}, \cB}_{\mathsf{r}}$ are given as
\begin{align*}
	\lambda^{\mathsf{p}, \cB}&=\frac{\sum_{i=0}^{\nB-k-1}k-\tau-1-i}{\nB-k}=\frac{1}{2}\bigg(3k-2\tau-\nB-1\bigg),\\
	\begin{split}
		C^{\mathsf{p}, \cB}_{\mathsf{r}}&=O\bigg(\frac{\sum_{i=0}^{\nB-k-1}k-\tau-2-i}{\nB-k}\nu\bigg)\\
		&\hspace{3.25cm}=O\bigg(\frac{1}{2}(3k-2\tau-\nB-3)\nu\bigg).
	\end{split}
\end{align*}

\subsection{Encoding Complexity}
\label{sec:EncComp}

The encoding complexity, denoted by $C_{\mathsf{e}}$, is the sum of the encoding complexities of Class $\cA$ and Class $\cB$ codes. The generation of each of the $\nA-k$ Class $\cA$ parity symbols in one row of the code array, $p^{\cA}_{i,j}$ in \eqref{eq:pij}, requires $k$ multiplications and $k-1$ additions. Adding data symbols to $\tau$ of these parity symbols according to \eqref{eq:piu} requires an additional $\tau$ additions. The encoding complexity of the Class $\cA$ code is therefore %given by 
\begin{align*}
	C^{\cA}_{\mathsf e}= O((\nA-k)(k\nu^2+(k-1)\nu))+ O(\tau\nu).
\end{align*}

According to Sections \ref{Sec:ClassB} and \ref{sec:improvement}, the parity symbols in the first Class $\cB$ parity node are constructed as  sums of at most $k-\tau-1$ data symbols, and each parity symbol %in the next class B  parity node is  constructed as  sums of at most $k-\tau-2$ data symbols, and so on.
 in the subsequent parity nodes is constructed as a sum of data symbols  from a set of size one less.   % uses one less data symbol. 
Therefore, the encoding complexity of the Class $\cB$ code is
	\begin{align} \label{eq:CB}
	\begin{split}
		C^{\cB}_{\mathsf e} &\leq \sum_{i=1}^{n-\nA} O((k-\tau-1-i)\nu)\\
		&= O\bigg(\frac{1}{2}(n-\nA)(2k-2\tau-3-n+\nA)\nu\bigg).
	\end{split}
%		\begin{cases}
%			O((k-\tau-2)\nu)+\Theta(\nu) & \text{if $k-\tau-1>1$} \\
%			O(\nu) & \text{if $k-\tau-1=1$}
%		\end{cases}
\end{align}
Note that for Construction~$1$ the upper bound on $C^{\cB}_{\mathsf e}$ in (\ref{eq:CB}) is tight.
%Hence the total encoding complexity per row in our proposed code array is 
Finally, $C_{\mathsf{e}}=C^{\cA}_{\mathsf e}+C^{\cB}_{\mathsf e}$.
%\begin{align}
%	C_{\textrm{E}}=\CA+\CB.
%\end{align}

%When one requires a higher rate code, puncturing is performed on Class $\cB$ parity nodes and thereby reduction in complexity. One can also see that in (4), the first term in $C_A$ is the largest term hence we can say that the encoding complexity per stripe is $O((\nA-k)(k\nu^2+(k-1)\nu))$.

\subsection{Code Comparison}

In this section, we compare the performance of the proposed codes with that of several codes in the literature, namely MDS codes, exact-repairable MDS codes \cite{Raw17}, MDR codes \cite{wan14}, Zigzag codes \cite{tam13}, Piggyback codes \cite{Ras17}, generalized Piggyback codes \cite{Yua16}, EVENODD codes \cite{Bla95}, Pyramid codes \cite{hua07}, and LRCs  \cite{hua12}. Throughout this section, we compare the repair bandwidth and the repair complexity of the systematic nodes with respect to other codes, except for exact-repairable MDS and BASIC PM-MBR codes. The reported repair bandwidth and complexity for these codes are for all nodes (both systematic and parity nodes).

\cref{Tab:Summary} provides a summary of the characteristics of the proposed codes as well as different codes proposed in the literature.\footnote{The variables $(t, t_r)$ and $r$ in \cref{Tab:Summary} are defined in \cite{Ras17} and \cite{hua12}, respectively. The definition of $\ell$ comes directly from $r$ that is defined in \cite{Ras17}.} In the table, column $2$ reports the value of $\beta$ (see \eqref{eq:lambda}) for each code construction. For our code, $\beta=k$, unlike for MDR and Zigzag codes, for which $\beta$ grows exponentially with $k$. This implies that our codes require less memory to cache data symbols during repair.
On the contrary, EVENODD codes have a lower sub-packetization and repair complexity, but this comes at the cost of having the same repair bandwidth as MDS codes.
The Piggyback codes presented in the table and throughout this section are from the piggybacking design $1$ in \cite{Ras17}, which provides efficient repair for only data nodes.
The fault tolerance $f$, the normalized repair bandwidth $\lambda$, the normalized repair complexity, and the encoding complexity, discussed in the previous subsections, are reported in columns $3$, $4$, $5$, and $6$, respectively.
\ifonecolumn
\begin{figure}[t]
	\centering
	\includegraphics{Comp1}
	\caption{Comparisons of different codes $(n,k,f)$ with $\nu=8$.} 
	%Repair complexity denotes the average number of operations performed to repair a failed data block}
	\label{fig5}
	\vspace{-3ex}
\end{figure}
\else
\begin{figure}[t]
	\centering
	\includegraphics[width=\columnwidth]{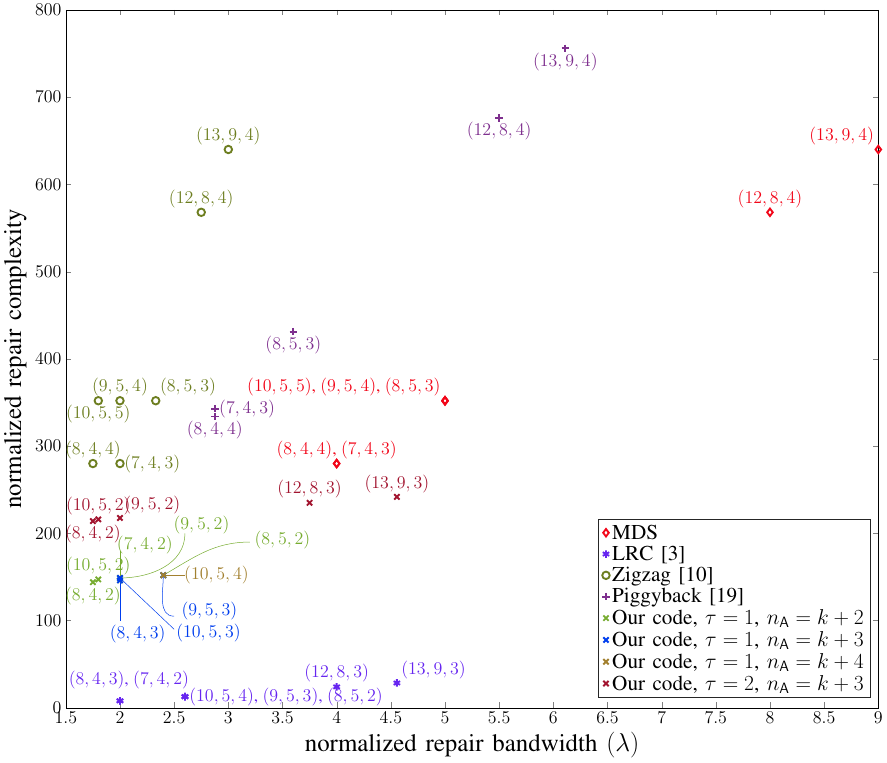}
	\caption{Comparisons of different $(n,k,f)$ codes with $\nu=8$.} 
	%Repair complexity denotes the average number of operations performed to repair a failed data block}
	\label{fig5}
	\vspace{-3ex}
\end{figure}
\fi
	
In \cref{fig5,Fig. CodeComp_p2}, we compare our codes with Construction~$1$ for the Class $\cB$ codes (i.e., the codes are constructed as shown in \cref{Sec:ClassA,Sec:ClassB}) with other codes in the literature. We remark that the Pyramid codes in \cref{Fig. CodeComp_p2} refer to the basic Pyramid codes in \cite{hua07}, while the exact-repairable MDS codes refer to the $(2,n-k,n-1)$ exact-repairable MDS codes from \cite[Sec.~IV]{Raw17}. The aforementioned notation, unlike our notation in this paper, refers to an $(n,k,n-k)$ code that has $\lambda\leq2\frac{n-1}{n-k}$, $\beta=n-k$, and repair locality of $n-1$. 
For generalized Piggyback codes \cite{Yua16}, we choose $\beta=k$. Also note that the parameters $s,p$ are chosen according to \cite[Eq.~20]{Yua16}, i.e., $s = \left\lfloor k \frac{\sqrt{n-k}-1}{\sqrt{n-k}} \right\rfloor$ or $s = \left \lceil k \frac{\sqrt{n-k}-1}{\sqrt{n-k}} \right\rceil$ and $p = k-s$, whichever pair of values gives the lowest repair bandwidth.  In case of a tie, the pair that gives the lowest repair complexity was chosen. 
%as equal as possible while keeping $s+p=\beta$.
In particular, the figure plots the normalized repair complexity of $(n, k, f)$ codes over $\text{GF}(2^8)$ ($\nu=8$) versus their normalized repair bandwidth $\lambda$. In the figure, we show the exact repair bandwidth for our proposed codes, while the reported repair complexities and the repair bandwidths of the other codes, except for Piggyback, generalized Piggyback, and exact-repairable MDS codes, are from \cref{Tab:Summary}.\footnote{For LRCs the expressions for the repair bandwidth and the repair complexity  tabulated in \cref{Tab:Summary} are used when $n-k-r$ is a divisor of $k$. When $n-k-r$ is not a divisor of $k$, exact values for the repair bandwidth and the repair complexity are calculated directly from the codes.} For Piggyback, generalized Piggyback, and exact-repairable MDS codes exact values for the repair bandwidth and the repair complexity are calculated directly from the codes. Furthermore, for a fair comparison we assume the parity symbols in the first parity node of all storage codes to be weighted sums. The only exception is the LRCs and the exact-repairable MDS codes, as the code design enforces the parity-check equations to be non-weighted sums. Thus, changing it would alter the maximum erasure correcting capability of the LRC and the repair bandwidth of the exact-repairable MDS code. We also assume that the LRCs and the Pyramid codes have a repair locality of $k/2$. For the generalized Piggyback codes, we assume that the codes have sub-packetization $\beta=k$. For the Piggyback codes, we consider the construction that repairs just the data nodes. Therefore, they have a sub-packetization of $2$.

%In contrast to the bounds for the repair bandwidth and complexity reported in \cref{Tab:Summary}, Fig.~\ref{fig5} contains the exact number of elementary binary additions.

\begin{table*}[!t]
     \caption{Comparison of normalized repair complexity and bandwidth of $(n,k,f)$ BASIC PM-MBR codes \cite{Hou16} and the proposed codes.}
    \label{Tab:RepairComplexity_compare}
    \centering
    \def\Hline{\noalign{\hrule height 2\arrayrulewidth}}
    \vskip -2.0ex %take space out between caption and table
    \begin{tabular}{@{}lcccccccccccc@{}}
        \Hline \\ [-2.0ex]
        \textbf{Proposed} & $n_\cA$ & $\tau$ & $R$ & $\mathrm{GF}(q)$ & \textbf{BASIC PM-MBR} &$R^{\mathsf b}$ &  $\delta^{\mathsf b}$  & $\mathcal R_m$ & $C_{\mathsf r}^{\mathsf b}$ & $C_{\mathsf r}$ & $\lambda^{\mathsf b}$ & $\lambda$  \\[0.25ex]
        \hline
        \\ [-2.0ex] \hline  \\ [-2.0ex]
        $(9,5,3)$ & $8$ & $1$& $0.5556$ & $\mathrm{GF}(11)$ &$(8,5,3)$ &0.4464 & $7$  & $\mathcal R_{11}$ & $135$ & $44$ & $1$ & $2.4$ \\
        $(11,7,3)$ & $10$ & $2$ &$0.6364$ & $\mathrm{GF}(11)$ & $(11,7,4)$ &$0.4454$ & $10$ & $\mathcal R_{11}$ & $187.5$ & $66.2857$ & $1$ & $3$ \\
        $(14,9,3)$ & $12$ & $2$& $0.6428$ & $\mathrm{GF}(13)$ & $(14,9,5)$ &$0.4450$ & $13$  & $\mathcal R_{17}$ & $384$ & $70.6667$ & $1$ & $3.5556$ \\
%        $(14,8)$ & $12$ & $3$ & $2.375$ & $2.3275$ \\
%        $(16,10)$ & $15$ & $4$ & $3.5$ & $3.45$ \\      
        \hline
    \end{tabular}
    %\vspace{-0.5ex}
    \vskip -2ex % take space after table
\end{table*}

The best codes for a DSS should be the ones that achieve the lowest repair bandwidth and have the lowest repair complexity. As seen in Fig.~\ref{fig5}, MDS codes have both high repair complexity and repair bandwidth, but they are optimal in terms of fault tolerance for a given $n$ and $k$. Zigzag codes achieve the same fault tolerance and high repair complexity as MDS codes, but at the lowest repair bandwidth. At the other end, LRCs yield the lowest repair complexity, but a higher repair bandwidth and worse fault tolerance than Zigzag codes. 
Piggyback codes, generalized Piggyback codes, and exact-repairable MDS codes have  a repair bandwidth between those of Zigzag and MDS codes, but with a higher repair complexity. Strictly speaking, they have a repair complexity higher than MDS codes.
%Piggyback codes have a repair bandwidth between those of Zigzag and MDS codes, but with a higher repair complexity. 
For a given storage overhead, our proposed codes have better repair bandwidth than MDS codes, Piggyback codes, generalized Piggyback codes, and exact-repairable MDS codes. In particular, the numerical results in \cref{fig5,Fig. CodeComp_p2} show that for different code parameters with $\tau=1$ and $\nA-k=2$ our proposed codes yield a reduction of the repair bandwidth in the range of $64\%-50\%$, $39.13\%-30.43\%$, $43.04\%-33.33\%$, and $33.33\%-30\%$, respectively. 
Furthermore, our proposed codes yield lower repair complexity as compared to MDS, Piggyback, generalized Piggyback, exact-repairable MDS, and Zigzag codes. Again, the numerical analysis in \cref{fig5,Fig. CodeComp_p2} shows that for different code parameters with $\tau=1$ and $\nA-k=2$ our proposed codes yield a reduction of the repair complexity in the range of $58.18\%-47.86\%$, $64.75\%-56.89\%$, $59.26\%-49.30\%$, $78.70\%-67.93\%$, and $58.18\%-47.86\%$, respectively. 
However, the benefits in terms of repair bandwidth and/or repair complexity with respect to MDS codes, Zigzag codes, and codes constructed using the piggybacking framework come at a price of a lower fault tolerance.
For fixed $(n,k,f)$ it can be seen that the proposed codes yield a reduction of the repair bandwidth in the range of $7.69\%-0\%$ compared to LRCs and Pyramid codes, while in some cases, for the latter codes our proposed codes achieve a reduction of the repair complexity in the range of $24.44\%-15.18\%$.

\ifonecolumn
\begin{figure}[t]
\centering
\includegraphics{Comp2}
\caption{Comparisons of different codes $(n,k,f)$ with $\nu=8$.}
\label{Fig. CodeComp_p2}	
\vskip -5ex
\end{figure}
\else
\begin{figure}[t]
\centering
\includegraphics[width=\columnwidth]{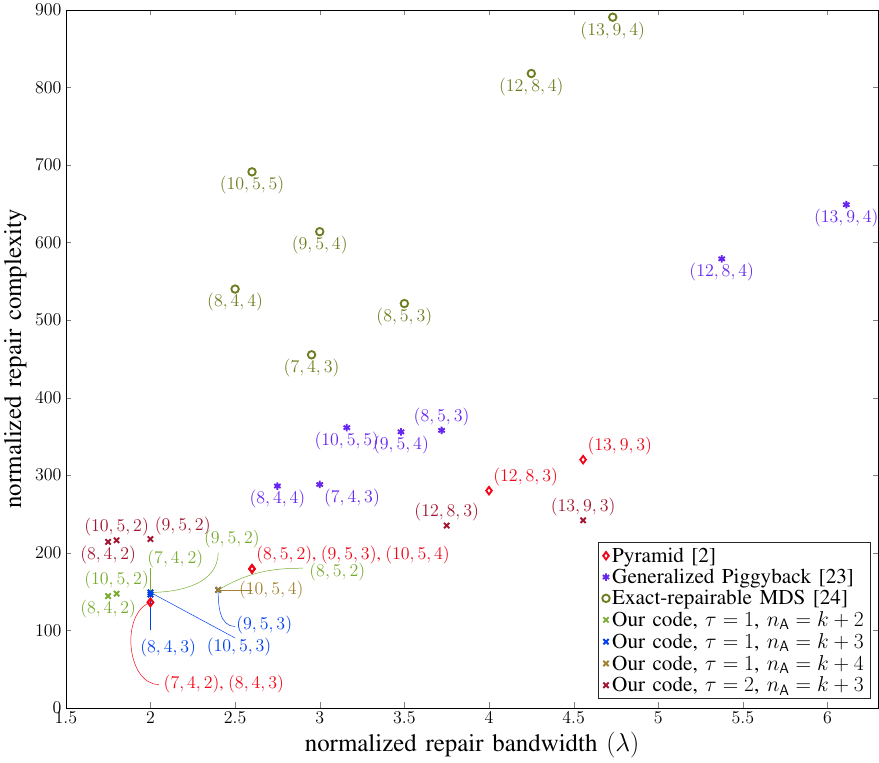}
\caption{Comparisons of different $(n,k,f)$ codes with $\nu=8$.}
\label{Fig. CodeComp_p2}	
\vskip -4ex
\end{figure}
\fi

In \cref{Tab:RepairComplexity_compare}, we compare the normalized repair complexity of the proposed codes with Construction~$1$ for the Class $\cB$ code and BASIC PM-MBR codes presented in \cite{Hou16}. BASIC PM-MBR codes are constructed from an algebraic ring $\mathcal R_m$, where each symbol in the ring is a binary vector of length $m$. In order to have a fair comparison with our codes, we take the smallest possible field size for our codes and the smallest possible ring size for the codes in \cite{Hou16}. Furthermore, to compare codes with similar storage overhead, we consider BASIC PM-MBR codes with repair locality, denoted by $\delta^{\mathsf b}$, that leads to the same file size as for our proposed codes (which is equal to $k^2$) and code length $n$ such that the code rate (which is the inverse of the storage overhead) is as close as possible to that of our proposed codes.
%Furthermore, for the BASIC PM-MBR codes, we vary the number of nodes contacted in repair in order to retain the same file size as for our code (which is equal to $k^2$).
%Furthermore, we consider that to repair a failed node $n-1$ nodes are contacted. 
The Class $\cA$ codes of the proposed codes in the table are $(n_\cA,n_\cA-f)$ RS codes. The codes are over $\text{GF}(n_\cA+1)$ if $n_\cA+1$ is a prime (or a power of a prime). If $n_\cA+1$ is not a prime (or a power of prime), we construct an $(n',n'-f)$ RS code over $\mathrm{GF}(n'+1)$, where $n'+1$ is the smallest prime (or the smallest power of a prime) with $n'+1\geq n_\cA+1$, and then we shorten the RS code to obtain an $(n_\cA,n_\cA-f)$ Class $\cA$ code. In the table, the parameters for the proposed codes are given in columns $1$ to $5$ and those of the codes in \cite{Hou16} in columns $6$ to $9$. The code rates $R$ and $R^{\mathsf b}$ of the proposed codes and the BASIC PM-MBR codes\footnote{BASIC PM-MBR codes have code rate $R^{\mathsf b}=B/n\delta^{\mathsf b}$, where $B=\binom{k+1}{2}+k(\delta^{\mathsf b}-k)$ is the file size and $\delta^{\mathsf b}\in\{k,k+1,\ldots,n-1\}$.} are given in columns $4$ and $7$, respectively. The smallest possible field size for our codes and the smallest possible ring size for the BASIC PM-MBR codes are given in columns $5$ and $9$, respectively. The normalized repair complexity\footnote{The normalized repair complexity of BASIC PM-MBR codes is $C^{\mathsf b}_{\mathsf r}=(3.5\delta^{\mathsf b}+2.5)(m-1)/2$ \cite{Hou16}. Such codes have $\beta=\delta^{\mathsf b}$, $\lambda=1$, and $f=n-k$. The value of $m$ is conditioned on the code length $n$ (see \cite[Th. 14]{Hou16}).} for BASIC PM-MBR codes, $C^{\mathsf b}_{\mathsf r}$, is given in column $10$, while the normalized repair complexity of the proposed codes, $C_{\mathsf r}$, is given in column $11$. The normalized repair bandwidth for BASIC PM-MBR codes, $\lambda^{\mathsf b}$, is given in column $12$, while the normalized repair bandwidth of the proposed codes, $\lambda$, is given in the last column. It can be seen that the proposed codes achieve significantly better repair complexity. However, this comes at the cost of a lower fault tolerance for the codes in rows $2$ and $3$ (but our codes have significantly higher code rate) and higher repair bandwidth (since BASIC PM-MBR codes are MBR codes, their normalized repair bandwidth is equal to $1$). For the $(9,5,3)$ code, the same fault tolerance of the $(8,5,3)$ code in \cite{Hou16}  is achieved, despite the fact that the proposed code has a higher code rate. We remark that FR codes achieve a better (trivial) repair complexity compared to our proposed codes. However, this comes at a cost of $R<0.5$ and they cannot be constructed for any  $n$, $k$, and $\delta$, where $\delta$ is the repair locality. 

%when the Class $\cB$ nodes are constructed as shown in \cref{Sec:ClassB,sec:improvement}. 

%\begin{table}[!t]
%     \caption{Comparison of normalized repair complexity of $(n,k,n-k)$ BASIC PM-MBR codes \cite{Hou16} and the $(n,k,f)$ proposed codes with $\tau=1$}
%    \label{Tab:RepairComplexity_compare}
%    \centering
%    \def\Hline{\noalign{\hrule height 2\arrayrulewidth}}
%    \vskip -2.0ex %take space out between caption and table
%    \begin{tabular}{@{}lcccccc@{}}
%        \Hline \\ [-2.0ex]
%        Code & $n_\cA$ &  $f$ &$\mathrm{GF}(q)$ & $\mathcal R_m$ & \textbf{BASIC PM-MBR} & \textbf{Proposed} \\[0.25ex]
%        \hline
%        \\ [-2.0ex] \hline  \\ [-2.0ex]
%        $(7,4)$ & $6$& $2$ & $\mathrm{GF}(7)$ & $\mathcal R_7$ & $70.5$ & $24.75$ \\
%        $(8,4)$ & $7$ & $3$ & $\mathrm{GF}(8)$ & $\mathcal R_{11}$ & $135$ & $24.75$ \\
%        $(9,5)$ & $8$ & $3$ & $\mathrm{GF}(11)$ & $\mathcal R_{11}$ & $152.5$ & $44$ \\
%        $(10,5)$ & $8$ & $3$ & $\mathrm{GF}(11)$ & $\mathcal R_{11}$ & $170$ & $42.4$ \\
%%        $(14,8)$ & $12$ & $3$ & $2.375$ & $2.3275$ \\
%%        $(16,10)$ & $15$ & $4$ & $3.5$ & $3.45$ \\      
%        \hline
%    \end{tabular}
%    %\vspace{-0.5ex}
%    %\vskip -2ex % take space after table
%\end{table}

In \cref{Tab:Heuristic_Comparision}, we compare the normalized repair bandwidth of the proposed codes using Construction~$1$ and Construction~$2$ for Class $\cB$ nodes. In the table, with $\lambda^{\mathsf{C}1}$ and $\lambda^{\mathsf{C}2}$, we refer to the normalized repair bandwidth for Construction~$1$ and Construction~$2$, respectively. For the codes presented, it is seen that the heuristic construction yields an improvement in repair bandwidth in the range of $1\%-7\%$ with respect to Construction~$1$.
\begin{table}[!t]
     \caption{Improvement in normalized repair bandwidth of the proposed $(n,k)$ codes when the Class $\cB$ nodes are heuristically constructed.}
    \label{Tab:Heuristic_Comparision}
    \centering
    \def\Hline{\noalign{\hrule height 2\arrayrulewidth}}
    \vskip -2.0ex %take space out between caption and table
    \begin{tabular}{@{}lccccc@{}}
        \Hline \\ [-2.0ex]
        Code & $\nA$ & $\tau$ & $\lambda^{{\mathsf C}1}$ & $\lambda^{{\mathsf C}2}$ & \textbf{Improvement} \\[0.25ex]
        \hline
        \\ [-2.0ex] \hline  \\ [-2.0ex]
        $(7,4)$ & $6$ & $1$ & $2$ & $1.875$ & $6.25\%$\\
        $(10,6)$ & $9$ & $2$ & $2.5$ & $2.4167$ & $3.33\%$\\
        $(13,8)$ & $12$ & $3$ & $3$ & $2.9375$ & $2.08\%$\\
        $(14,8)$ & $12$ & $3$ & $2.375$ & $2.3125$ & $2.63\%$\\
        $(16,10)$ & $15$ & $4$ & $3.5$ & $3.45$ & $1.43\%$\\      
        \hline
    \end{tabular}
    %\vspace{-0.5ex}
\vskip -5ex % take space after table
\end{table}

%\vspace{-0.1cm}
\section{Conclusion} \label{sec:conclu}

In this paper, we constructed a new class of codes that achieve low repair bandwidth and low repair complexity for a single node failure. The codes are constructed from two smaller codes, Class $\cA$ and $\cB$, where the former focuses on the fault tolerance of the code, and the latter focuses on reducing the repair bandwidth and complexity. 
It is numerically seen that our proposed codes achieve better repair complexity than Zigzag codes, MDS codes, Piggyback codes, generalized Piggyback codes, exact-repairable MDS codes, BASIC PM-MBR codes, and are in some cases better than Pyramid codes. They also achieve a better repair bandwidth compared to MDS codes, Piggyback codes, generalized Piggyback codes, exact-repairable MDS codes, and are in some cases better than LRCs and Pyramid codes.
%Our proposed codes achieve better repair complexity than Zigzag codes, Piggyback codes, and BASIC PM-MBR codes and better repair bandwidth than LRCs, at the cost of slightly lower fault tolerance. 
A side effect of such a construction is that the number of symbols per node that need to be encoded grows only linearly with the code dimension. This implies that our codes are suitable for memory constrained DSSs as compared to Zigzag and MDR codes, for which the number of symbols per node increases exponentially with the code dimension. %Furthermore, we present an heuristic algorithm to construct the Class $\cB$ nodes, which gives an additional improvement in the repair bandwidth when the code parameter $k$ is even.

\appendices
\section{Proof of \cref{th:ECC}}
\label{App: Proof_of_fault_tolerance}
%We start with the following observation.
Consider an arbitrary set of $\tau'\leq \tau$ piggybacked nodes, denoted by $\mathcal T=\{j_1,j_2,\ldots,j_{\tau'}\}$,  $j_{i}=j'_{i}-(\nA-\tau)+1$, $j'_{i}\in\{\nA-\tau,\ldots,\nA-1\}$. Then, the repair of the $i$-th row, $i=0,\ldots,k-1$, using the piggybacked nodes in $\mathcal T$, would depend upon the knowledge of the  data symbols (piggybacks) in the rows $(i+j_1)_k,(i+j_2)_k,\ldots,(i+j_{\tau'})_k$. This is because the knowledge of the piggybacks in these rows allows to obtain the original MDS parity symbols in the $i$-th row. In the following, we use this observation. 
We first proceed to prove that  if $\tau'(\tau'+\nA-k-\tau)<k$, then $\theta+\tau'\leq\nA-k-\tau+\tau'$ data nodes can be corrected using $\theta$ non-modified parity nodes and the $\tau'$ piggybacked nodes in $\mathcal T$. 
Using this we will complete the proof of \cref{th:ECC}.
\begin{lemma}
\label{lem:AppA}
	Consider an $(\nA,k)$ Class $\cA$ code with $k+2\leq\nA<2k$. The code consists of $\tau$ piggybacked nodes and $\nA-k-\tau$ non-modified MDS parity nodes.  If $\tau'(\tau'+\nA-k-\tau)<k$,  then the code can correct $\theta+\zeta$ data node failures using $\theta\leq\nA-k-\tau$ non-modified parity nodes and the $\tau'$ piggybacked parity nodes in the set $\mathcal T=\{1,\ldots,\tau'\}$ for $\zeta\leq\tau'\leq\tau$.
\end{lemma}
\begin{IEEEproof}
We consider first the case when $\zeta=\tau'$. Then, assume that $\theta+\zeta$ data nodes fail and there exists a sequence $i, (i+1)_k,\ldots,(i-1+\tau')_k$ of $\tau'$ data nodes that are available. By construction, the parity symbol $p^{\cA,\mathsf p}_{(j-1)_k,\nA-\tau-1+t}$, $t\in\mathcal T$, is (see \eqref{eq:piu}) given by
\begin{align}
\label{eq:piggyback_lemma}
p^{\cA,\mathsf p}_{(j-1)_k,\nA-\tau-1+t}= p^{\cA}_{(j-1)_k,\nA-\tau-1+t}  + d_{(j-1+t)_k, (j-1)_k},
\end{align}
where $j=0,\ldots,k-1$. 
%Let $i$ be the largest row index such that the symbol $d_{(i-1+t)_k,(i-1)_k}$ is not erased (such a non-erased data symbol always exists since not all data nodes fail). 
To recover all data symbols, set $i'=(i+\tau')_k$ and perform the following steps. 
\begin{enumerate}
\item Obtain the $\zeta=\tau'$ MDS parity symbols $p^\cA_{(i'-1)_k,\nA-\tau-1+t}$, $t \in \mathcal{T}$ (see \eqref{eq:piggyback_lemma}) in the $(i'-1)_k$-th row. This is possible because the piggybacks in the $(i'-1)_k$-th node are available.
%\item Obtain the MDS parity symbol $p^\cA_{(i'-1)_k,(\nA-\tau)}$ (see \eqref{eq:piggyback_lemma}) in the $(i'-1)_k$-th row.
\item Using the $\theta+\zeta$ MDS parity symbols and $k-\theta-\zeta$ data symbols in the $(i'-1)_k$-th row, recover the missing $\theta+\zeta$  symbols in the $(i'-1)_k$-th row of the failed data nodes.
\item $i'\leftarrow (i'-1)_k$.
%\item , from which the piggyback symbol $d_{(i'-1)_k,(i'-2)_k}$ is obtained. 
\item Repeat Items 1), 2), and 3) $\tau'-1$ times.
This ensures that the failed symbols in the $\tau'$ rows $i',(i'+1)_k,\ldots,(i'-1+\tau')_k$ are recovered. This implies that the piggyback symbols 
\begin{align}
\label{Eq: Piggybacks_recovered}
	\begin{split}
		&d_{i',(i'-1)_k},\ldots,d_{(i'-1+\tau')_k,(i'-1)_k}\,\, \text{in node}\,\, (i'-1)_k,\\ 
		&d_{i',(i'-2)_k},\ldots,d_{(i'-2+\tau')_k,(i'-2)_k}\,\, \text{in node}\,\, (i'-2)_k,\\ 
		&\,\,\,\,\,\,\,\,\,\,\,\,\,\,\,\,\,\,\,\,\,\,\,\,\,\,\,\,\,\vdots\\
		&d_{i',(i'-\tau')_k}\,\,\,\,\,\,\,\,\,\,\,\,\,\,\,\,\,\,\,\,\,\,\,\,\,\,\,\,\,\,\,\,\,\,\,\,\,\,\,\,\,\,\,\,\,\,\,\,\,\,\,\,\,\,\,\, \text{in node}\,\, (i'-\tau')_k,
	\end{split}
\end{align}
are recovered. In other words, \cref{Eq: Piggybacks_recovered} says that in the $(i'-t)_k$-th node, $\tau'+1-t$ piggybacked symbols are recovered. More specifically, for $t=1$, $\tau'$ piggybacked symbols are recovered. Thus,
	\item  repeat Items 1) and 2), and set
	\item $i'\leftarrow (i'-1)_k$.
We thus recover the $i'$-th row and obtain the piggyback symbols in $\mathcal D_{\mathcal R_{i'}}\backslash\mathcal D_{\mathcal X_{i'}}$. This increases the number of obtained piggyback symbols by $1$ in the next $\tau'$ nodes $i',(i'-1)_k,\ldots,(i'+\tau'-1)_k$. In a similar fashion, we now have $\tau'$ piggybacked symbols in the $i'$-th node, and Items 5) and 6) are repeated until all $k$ rows have been recovered. With this recursion, one recovers the $\theta+\tau'=\theta+\zeta$ failed data nodes. 
\end{enumerate}
For the case when $\zeta<\tau'$, the aforementioned decoding procedure is still able recover $\theta+\zeta$ data node failures. This is because, in order to repair failed symbols in the $(i'-1)_k$-th row, one needs just $\zeta<\tau'$ MDS parity symbols from $\mathcal T'\subseteq\mathcal T$. If the $(i'-1)_k$-th node is available, then $\zeta$ piggybacks allow to recover the MDS parity symbols (see Item 1)). On the contrary, if the $(i-1)_k$-th node has been erased, then $\zeta$ piggybacks are obtained from \cref{Eq: Piggybacks_recovered}.

Note that the argument above assumes that $\tau'$ consecutive data nodes are available. Thus, in order to guarantee that any  $\theta+\zeta$ data nodes can be corrected, we consider the worst case scenario for node failures, where we equally spread $\theta+\zeta$ data node failures across the $k$ data nodes. Since
\begin{align*}
	\frac{k}{\theta+\zeta}\geq\frac{k}{\nA-k-\tau+\tau'}>\tau',
\end{align*}
where the last inequality follows by the assumption on $\tau'$ stated in the lemma, it follows that the largest gap of non-failed data nodes in the worst case scenario is indeed greater than or equal to $\tau'$.
\end{IEEEproof}

The above lemma shows that the Class $\cA$ code can correct up to $\nA-k-\tau+\tau'$ erasures using non-modified parity nodes and $\tau'$ modified parity nodes, provided the condition on $\tau'$ is satisfied. 

To prove that the code can correct $\nA-k-\tau+\tau'$ arbitrary node failures, let us assume that $\rho$ Class $\cA$ parity nodes and $\nA-k-\tau+\tau'-\rho$ data nodes have failed. More precisely, let $\rho_1\leq\nA-k-\tau$ non-modified nodes, $\rho_2\leq\tau'$ piggybacked nodes in $\mathcal T$, and $\rho_3\geq0$ remaining piggybacked nodes, where $\rho_1+\rho_2+\rho_3=\rho$, fail.  Clearly, it can be seen that there are $\nA-k-\tau-\rho_1$ non-modified parity nodes and  a set of $\zeta=\tau'-\rho_2$ modified nodes $\mathcal T'\subseteq\mathcal T$ available. Also, note that the number of data node failures is $\rho_3$ less than the number of  combined available piggybacked nodes in $\mathcal T'$ and available non-modified parity nodes. Thus,  using \cref{lem:AppA} with $\theta=\nA-k-\tau-\rho_1-\rho_3$ and $\zeta=\tau'-\rho_2$, it follows that $\theta+\zeta=(\nA-k-\tau-\rho_1-\rho_3)+(\tau'-\rho_2) = \nA-k-\tau+\tau'-\rho$ data nodes can be repaired. The remaining $\rho$ failed parity nodes can then be repaired using the $k^2$ data symbols in the $k$ data nodes.

We remark that the decoding procedure in \cref{lem:AppA} in essence solves a system of linear equations by eliminating $\tau' k$ variables (piggybacks) in $\tau'$ parity nodes. Once the piggybacks are eliminated, the $k(\theta+\tau')$ data symbols are obtained by solving $k$ systems of linear equations. Thus, the decoding procedure is optimal, i.e., it is maximum likelihood decoding. 

Consider the quadratic function  $\psi(\tau')={\tau'}^2+(\nA-k-\tau)\tau'-k$. According to the proof of \cref{lem:AppA} when $\psi(\tau')\geq0$, the decoding procedure fails as  one can construct a failure pattern for the data nodes where the largest separation (in the number of available nodes) between the failed nodes would be strictly smaller than $\tau'$. The largest $\tau'$ such that $\psi(\tau')<0$ can be determined as follows. By simple arithmetic, one can prove that $\psi(\tau')$ is a convex function with a  negative minima and with a positive and a negative root. Therefore,
\begin{align*}
	0\leq\tau'<\xi=\frac{\sqrt{(\nA-k-\tau)^2+4k}-(\nA-k-\tau)}{2},
\end{align*} 
where $\xi$ is the positive root of $\psi(\tau')$. Furthermore, it may happen that $\tau<\xi$. Therefore, the maximum number of node failures that the code can tolerate is
\begin{align*}
	f=\begin{cases}
	 \nA-k-\tau+\bigg\lfloor\frac{\sqrt{(\nA-k-\tau)^2+4k}-(\nA-k-\tau)}{2}\bigg\rfloor & \text{if}\,\, \tau\geq\xi\\
	 \nA-k & \text{if}\,\, \tau<\xi	
	\end{cases}.
\end{align*}

\section{Class $\cB$ Parity Node Construction}
\label{App: Algorithm}

\begin{algorithm}[t]
\SetArgSty{textup}
\SetKwFunction{ConstNode}{ConstructNode}
\SetKwFunction{ConstLastNode}{ConstructLastNode}
\SetKwFunction{UpdateA}{UpdateReadCost}
\SetKwInput{Init}{Initialization}
\Init{\\ $\bm A=[a_{i,j}]$ as defined in (\ref{Eq:A})
\\ $n=\nA+\nB-k$ with    $\nB < 2k-\tau$ and $\nA < 2k$ %$n< k-\tau-1+\nA$ %$\forall i,j=0,\ldots,k-1$
%		  \\ $\mathcal{\tilde{D}}_{i,j}\triangleq\{d_{(i+s)_k,(j+s)_k}\}_{s=0}^{k-1}$ 
%		  \\ \Indm $max\_itr=k-\tau-2$
		  }
\For{$l \in \{\nA,\ldots,n-1\}$}{
$\rho_l \leftarrow k-\tau-1-l+\nA$\\
\uIf{$l\leq\nA+k/2-\tau-2$}{
$\mathcal P^{\cB,\mathsf{h}}_l \leftarrow \mathcal P^{\cB}_l$ \label{alg:NodeConstruction_no_change}
}\uElseIf{$l>\nA+k/2-\tau-2$ and $\rho_l>1$} {
$\mathcal P^{\cB,\mathsf{h}}_l \leftarrow $ \ConstNode{$\bm A,\rho_l$}
}\ElseIf{$l>\nA+k/2-\tau-2$ and $\rho_l=1$}{
$\mathcal P^{\cB,\mathsf{h}}_l \leftarrow $ \ConstLastNode{}
}
$\bm A \leftarrow $ \UpdateA{}
}
\caption{Class $\cB$ node construction when $k$ is even}\label{alg:NodeConstruction}	
\end{algorithm}

In this appendix, we give an algorithm that constructs $\nB-k$ Class $\cB$ parity nodes $\mathcal P^{\cB,\mathsf{h}}_{\nA},\ldots,\mathcal P^{\cB,\mathsf{h}}_{n-1}$. The algorithm is a heuristic for the  construction of Class $\cB$ parity nodes such that the repair bandwidth of failed nodes is further reduced in comparison with Construction~$1$ in \cref{Sec:ClassB}.

The nodes $\mathcal P^{\cB,\mathsf{h}}_l$, $l=\nA,\ldots,n-1$, are constructed recursively as shown in \cref{alg:NodeConstruction}. The $k$ parity symbols in the $l$-th node are sums of at most $\rho_l$ data symbols $d_{i,j}\in\cup_{j'}\mathcal D_{\mathcal Q_{j'}}$. The construction of the $l$-th node for $l\leq\nA+k/2-\tau-2$ (see \cref{alg:NodeConstruction_no_change}) is identical to that resulting from Construction~$1$ in \cref{Sec:ClassB}. The remaining parity nodes $\mathcal P^{\cB,\mathsf{h}}_l$, $l>\nA+k/2-\tau-2$, are constructed using the sub-procedures \texttt{ConstructNode} ($\bm A, \rho_l$) and \texttt{ConstructLastNode} ($\,\,$). After the construction of each parity node, the read costs of the data symbols $d_{i,j}$ are updated by the sub-procedure \texttt{UpdateReadCost} ($\,\,$).  In the following, we describe each of the above-mentioned sub-procedures.

\subsection{\texttt{ConstructNode} ($\bm A,\rho_l$)} 
\label{App: Subsection_algo}
This sub-procedure allows the construction of the $l$-th Class $\cB$ parity node, where each parity symbol in the node is a  sum of at most $\rho_l$ data symbols. The algorithm for the sub-procedure is shown in \cref{alg:SeedConstruction}. Here, the algorithm is divided into two parts. The first part (\cref{alg:SeedConstruction_P1s} to \cref{alg:SeedConstruction_P1e}) adds at most two data symbols to each of the $k$ parity symbols, while the second part (\cref{alg:SeedConstruction_P2s} to \cref{alg:SeedConstruction_P2e}) adds at most $\rho_l-2$ data symbols.

In the first part, each parity symbol $p^\cB_{t,l}$, $t=0,\ldots,k-1$, is recursively constructed by adding a symbol $d_{i,j}\in\mathcal D_{\mathcal{Q}_j\backslash\cup_{j'}\mathcal U_{j'}}$ which  has a corresponding read cost $a_{i,j}$ that is the largest among all symbols indexed by  $\mathcal{Q}_j\backslash\cup_{j'}\mathcal U_{j'}$ (see \cref{alg:SeedConstruction_Choose_ii} and \cref{alg:SeedConstruction_Choose_ij}). The next symbol added to $p^\cB_{t,l}$ is $d_{j,i}\in\mathcal D_{\mathcal X_j}$ if such a symbol exists (\cref{alg:SeedConstruction_Choose_ji}). Otherwise, the symbol added is $d_{j,i_2}\in\mathcal D_{\mathcal X_j\backslash\cup_{j'}\mathcal U_{j'}}$ if such a symbol exists (\cref{alg:SeedConstruction_Choose_jiPP}). The set $\mathcal U_{t}$ denotes the index set of data symbols from $\mathcal D$ that are added to $p^\cB_{t,l}$. Note that there exist multiple choices for the symbol $d_{i,j}$. A symbol $d_{i,j}$ such that there is a valid $d_{j,i}\in\mathcal D_{\mathcal X_j}$ is preferred, since it allows a larger reduction of the repair bandwidth.

The second part of the algorithm chooses recursively at most $\rho_l-2$ data symbols that should participate in the construction of the $k$ parity symbols. The algorithm chooses a symbol $d_{j,i_3}\in\mathcal D_{\mathcal X_j\backslash\cup_{j'}\mathcal U_{j'}}\not=\emptyset$ such that $\readd(d_{j,i_3},p^\cB_{t,l}+d_{j,i_3})< a_{j,i_3}$ (see \cref{alg:SeedConstruction_Choose_ji3}). In other words, choose data symbols such that their read cost do not increase. It may happen that $\mathcal D_{\mathcal X_j\backslash\cup_{j'}\mathcal U_{j'}}=\emptyset$. If so, select $d_{j_2,i}\in\cup_{j_1}\mathcal D_{\mathcal X_{j_1}\backslash\cup_{j_1'}\mathcal U_{j_1'}}$ such that  $d_{i',j_2}\in\mathcal D_{\mathcal U_{t}}$ and  $a_{j_2,i}>1$, for some $i' \neq i$, exists, and then add $d_{j_2,i}$ to the parity symbol $p^\cB_{t,l}$ (see \cref{alg:SeedConstruction_Choose_j2i}). If $d_{j_2,i}$ does not exist, then an arbitrary symbol $d_{j_3,i}\in\cup_{j_1}\mathcal D_{\mathcal X_{j_1}\backslash\cup_{j_1'}\mathcal U_{j_1'}}$ is added (see \cref{alg:SeedConstruction_Choose_j3i}). This process is then repeated $\rho_l-2$ times.

\ifonecolumn
{
%\begin{tiny}
\begin{algorithm}[h]
\algsetup{linenosize=\tiny}
  \footnotesize
\SetKwBlock{Begin}{Begin}{}
\SetArgSty{textup}
\SetKwInput{Init}{Initialization}
\Init{\\ \Indp $t,j\leftarrow0$
	       \\  $\mathcal U_{t_1} \leftarrow \emptyset,\,\,\,\,\,t_1=0,\ldots,k-1$
	        %as defined in Appendix~\ref{App: Algorithm}
		  }
\While{$\argmin_{\forall t_1} |\mathcal U_{t_1}|< 2$}{\label{alg:SeedConstruction_P1s}
Select $d_{i,j}\in \mathcal D_{\Psi(\bm A_{\mathcal Q_j\backslash\cup_{j'}\mathcal U_{j'}})}$ s.t. $\exists\, d_{j,i}\in\mathcal D_{\mathcal X_j},a_{j,i}>1$\label{alg:SeedConstruction_Choose_ii}\\
\eIf{$d_{i,j}$ \emph{exists}}{
$p^\cB_{t,l} \leftarrow d_{i,j}+d_{j,i}$; $t\leftarrow t+1$;\label{alg:SeedConstruction_Choose_ji}\\ 
$\mathcal U_{t}\leftarrow \mathcal U_{t} \cup \{(i,j),(j,i)\}$ 
}{
Select $d_{i_1,j}\in\mathcal D_{\Psi(\bm A_{\mathcal Q_j\backslash\cup_{j'}\mathcal U_{j'}})}$\\ \label{alg:SeedConstruction_Choose_ij}
	\If{$d_{i_1,j}$ \emph{exists}}{
	\eIf{$\exists\, d_{j,i_2}\in\mathcal D_{\mathcal X_j \backslash\cup_{j'}\mathcal U_{j'}}$ s.t.\ $\readd(d_{j,i_2},p^\cB_{t,l}+d_{j,i_2})< a_{j,i_2}$ and $a_{j,i_2}>1$}{
	$p^\cB_{t,l} \leftarrow d_{i_1,j}+d_{j,i_2}$;\label{alg:SeedConstruction_Choose_jiPP}\\ $t\leftarrow t+1$; $\mathcal U_{t}\leftarrow \mathcal U_{t} \cup \{(i_1,j),(j,i_2)\}$ 
}{
$p^\cB_{t,l} \leftarrow d_{i_1,j}$; $t\leftarrow t+1$;\\
$\mathcal U_t\leftarrow\mathcal U_t\cup\{(i_1,j),(-1,-1)\}$
}
	}
}
$j\leftarrow(j+1)_k$
}\label{alg:SeedConstruction_P1e}
\caption{\texttt{ConstructNode} $(\bm A,\rho_l)$ Part 1/2}\label{alg:SeedConstruction}
\end{algorithm}
\setcounter{algocf}{1}
\begin{algorithm}
\algsetup{linenosize=\tiny}
  \footnotesize
\setcounter{AlgoLine}{19}
$t,j\leftarrow 0$\\
\While{$\argmin_{\forall t_1} |\mathcal U_{t_1}|\leq\rho_l$}{\label{alg:SeedConstruction_P2s}
\eIf{$\mathcal D_{\mathcal X_j\backslash \cup_{j_1}\mathcal U_{j_1} }\not=\emptyset$}{
%$\hat{\mathcal D}_{t}\leftarrow\{\hat{\mathcal D}_{t},dummy\_var\}$\\
	\uIf{$\exists\, d_{j,i_3}\in\mathcal D_{\mathcal X_j\backslash\cup_{j'}\mathcal U_{j'}}$ s.t.\ $\readd(d_{j,i_3},p^\cB_{t,l}+d_{j,i_3})< a_{j,i_3}$ and $a_{j,i_3}>1$}{\label{alg:SeedConstruction_Choose_ji3}
	$p^\cB_{t,l} \leftarrow p^\cB_{t,l}+d_{j,i_3}$; $t\leftarrow t+1$;\\ $\mathcal U_{t}\leftarrow\mathcal U_{t} \cup \{(j,i_3)\}$ 
	}\ElseIf{$l>\nA$}{
		$\mathcal U_{t}\leftarrow \mathcal U_{t} \cup (-1,-1)\}$; $t\leftarrow t+1$
	}
}{
	\If{$|\mathcal U_{t}|\leq \rho_l-1$}{
	Select $d_{j_2,i}\in\cup_{j_1}\mathcal D_{\mathcal X_{j_1}\backslash\cup_{j_1'}\mathcal U_{j_1'}}$ s.t. $d_{i',j_2}\in\mathcal D_{\mathcal U_{t}}$ for some $i' \neq i$\\
		\eIf{$d_{j_2,i}$ \emph{exists}}{
		$p^\cB_{t,l} \leftarrow p^\cB_{t,l}+d_{j_2,i}$; $t\leftarrow t+1$;\label{alg:SeedConstruction_Choose_j2i}\\ $\mathcal U_{t}\leftarrow \mathcal U_{t} \cup \{(j_2,i)\}$ 
		}{
		Select $d_{j_3,i}\in\cup_{j_1}\mathcal D_{\mathcal X_{j_1}\backslash\cup_{j_1'}\mathcal U_{j_1'}}$\\
		$p^\cB_{t,l} \leftarrow p^\cB_{t,l}+d_{j_3,i}$; $t\leftarrow t+1$;\label{alg:SeedConstruction_Choose_j3i}\\ $\mathcal U_{t}\leftarrow \mathcal U_{t} \cup \{(j_3,i)\}$ 
		}
	}
}
$j\leftarrow(j+1)_k$
}\label{alg:SeedConstruction_P2e}
\KwRet{$\{p^\cB_{0,l},\ldots,p^\cB_{k-1,l}\}$}
\caption{\texttt{ConstructNode} $(\bm A,\rho_l)$ Part 2/2}	
\end{algorithm}
\else
\begin{algorithm}[h]
\SetArgSty{textup}
\SetKwInput{Init}{Initialization}
\Init{\\ %\Indp 
$t,j\leftarrow0$
	       \\  $\mathcal U_{t_1} \leftarrow \emptyset,\,\,\,\,\,t_1=0,\ldots,k-1$
	        %as defined in Appendix~\ref{App: Algorithm}
		  }
\While{$\min_{\forall t_1} |\mathcal U_{t_1}|< 2$}{\label{alg:SeedConstruction_P1s}
Select $d_{i,j}\in \mathcal D_{\Psi(\bm A_{\mathcal Q_j\backslash\cup_{j'}\mathcal U_{j'}})}$ s.t. $\exists\, d_{j,i}\in\mathcal D_{\mathcal X_j},a_{j,i}>1$\label{alg:SeedConstruction_Choose_ii}\\
\eIf{$d_{i,j}$ \emph{exists}}{
$p^\cB_{t,l} \leftarrow d_{i,j}+d_{j,i}$ \label{alg:SeedConstruction_Choose_ji}\\ 
$\mathcal U_{t}\leftarrow \mathcal U_{t} \cup \{(i,j),(j,i)\}$; $t\leftarrow t+1$ 
}{
Select $d_{i_1,j}\in\mathcal D_{\Psi(\bm A_{\mathcal Q_j\backslash\cup_{j'}\mathcal U_{j'}})}$\\ \label{alg:SeedConstruction_Choose_ij}
	\If{$d_{i_1,j}$ \emph{exists}}{
	\eIf{$\exists\, d_{j,i_2}\in\mathcal D_{\mathcal X_j \backslash\cup_{j'}\mathcal U_{j'}}$ s.t.\ $\readd(d_{j,i_2},p^\cB_{t,l}+d_{j,i_2})< a_{j,i_2}$ and $a_{j,i_2}>1$}{
	$p^\cB_{t,l} \leftarrow d_{i_1,j}+d_{j,i_2}$\label{alg:SeedConstruction_Choose_jiPP}\\  $\mathcal U_{t}\leftarrow \mathcal U_{t} \cup \{(i_1,j),(j,i_2)\}$; $t\leftarrow t+1$ 
}{
$p^\cB_{t,l} \leftarrow d_{i_1,j}$\\
$\mathcal U_t\leftarrow\mathcal U_t\cup\{(i_1,j),(-1,-1)\}$; $t\leftarrow t+1$
}
	}
}
$j\leftarrow(j+1)_k$
}\label{alg:SeedConstruction_P1e}
$t,j\leftarrow 0$\\
\While{$\min_{\forall t_1} |\mathcal U_{t_1}|\leq\rho_l$}{\label{alg:SeedConstruction_P2s}
\eIf{$\mathcal D_{\mathcal X_j\backslash \cup_{j_1}\mathcal U_{j_1} }\not=\emptyset$}{
%$\hat{\mathcal D}_{t}\leftarrow\{\hat{\mathcal D}_{t},dummy\_var\}$\\
	\uIf{$\exists\, d_{j,i_3}\in\mathcal D_{\mathcal X_j\backslash\cup_{j'}\mathcal U_{j'}}$ s.t.\ $\readd(d_{j,i_3},p^\cB_{t,l}+d_{j,i_3})< a_{j,i_3}$ and $a_{j,i_3}>1$}{\label{alg:SeedConstruction_Choose_ji3}
	$p^\cB_{t,l} \leftarrow p^\cB_{t,l}+d_{j,i_3}$\\ $\mathcal U_{t}\leftarrow\mathcal U_{t} \cup \{(j,i_3)\}$; $t\leftarrow t+1$ 
	}\ElseIf{$l>\nA$}{
		$\mathcal U_{t}\leftarrow \mathcal U_{t} \cup (-1,-1)\}$; $t\leftarrow t+1$
	}
}{
	\If{$|\mathcal U_{t}|\leq \rho_l-1$}{
	Select $d_{j_2,i}\in\cup_{j_1}\mathcal D_{\mathcal X_{j_1}\backslash\cup_{j_1'}\mathcal U_{j_1'}}$ s.t. $d_{i',j_2}\in\mathcal D_{\mathcal U_{t}}$ and $a_{j_2,i}>1$ for some $i' \neq i$\\
		\eIf{$d_{j_2,i}$ \emph{exists}}{
		$p^\cB_{t,l} \leftarrow p^\cB_{t,l}+d_{j_2,i}$\label{alg:SeedConstruction_Choose_j2i}\\ $\mathcal U_{t}\leftarrow \mathcal U_{t} \cup \{(j_2,i)\}$; $t\leftarrow t+1$ 
		}{
		Select $d_{j_3,i}\in\cup_{j_1}\mathcal D_{\mathcal X_{j_1}\backslash\cup_{j_1'}\mathcal U_{j_1'}}$\\
		$p^\cB_{t,l} \leftarrow p^\cB_{t,l}+d_{j_3,i}$\label{alg:SeedConstruction_Choose_j3i}\\ $\mathcal U_{t}\leftarrow \mathcal U_{t} \cup \{(j_3,i)\}$; $t\leftarrow t+1$
		}
	}
}
$j\leftarrow(j+1)_k$
}\label{alg:SeedConstruction_P2e}
\KwRet{$\{p^\cB_{0,l},\ldots,p^\cB_{k-1,l}\}$}
\caption{\texttt{ConstructNode} $(\bm A,\rho_l)$}
\label{alg:SeedConstruction}
\end{algorithm}
%%\end{tiny}
%}
\fi

\subsection{\texttt{ConstructLastNode} ($\,\,$)}

This procedure constructs the $l$-th Class $\cB$ parity node that has $\rho_l=1$. In other words, each parity symbol $p^\cB_{t,l}$ is a data symbol $d_{i,j}\in\cup_{j'}\mathcal D_{\mathcal Q_{j'}}$. %The procedure is a two step procedure as follows. 
The procedure works as follows. 
First, initialize $\mathcal{U}_t$ to be the empty set for $t=0,\ldots,k-1$. Then, for $t=0,\ldots,k-1$, assign first the data symbol $d_{i,j}$ to the  parity symbol $p^\cB_{t,l}$, where  $d_{i,j} \in\mathcal D_{\Psi(\bm A_{\cup_{j'}\mathcal Q_{j'} \setminus  \cup_{t'} \mathcal{U}_{t'}})}$, and then subsequently add $(i,j)$ to $\mathcal{U}_{t}$. 
%\begin{align*}
	%p^\cB_{t,l} =  d_{i,j}=f^{-1}\bigg(\max_{d_{i',j'}\in\cup_{j'}\mathcal Q_{j'}} a_{i',j'}\bigg),
%	p^\cB_{t,l} =  \argmax_{d \in\cup_{j'}\mathcal Q_{j'}} f^{-1}(d),
%$p^\cB_{t,l} = d_{i,j}$ where  $d_{i,j} \in\mathcal D_{\Psi(\bm A_{\cup_{j'}\mathcal Q_{j'} \setminus  \cup_{t'} \mathcal{U}_{t'}})}$. Third, add $($
%\end{align*}
%In the last step, $(i,j)$ is added to $\mathcal{U}_{t}$.  
Note that for each iteration there may exist several choices for $d_{i,j}$, in which case we can pick one of these randomly.

\subsection{\texttt{UpdateReadCost} ($\,\,$)}
After the construction of the $l$-th node, we update the read costs of all data symbols $d_{i,j}\in\cup_{j'}\mathcal D_{\mathcal Q_{j'}}$. These updated values are used during the construction of the $(l+1)$-th node. The read cost updates for the parity symbol $p^\cB_{t,l}$, $t=0,\ldots,k-1$,  are
\begin{align*}
	a_{i,j}=\readd(d_{i,j},p^\cB_{t,l}),\,\,\,\forall d_{i,j}\in\mathcal D_{\mathcal U_t}.
\end{align*}

\ifCLASSOPTIONcaptionsoff
  \newpage
\fi

%\newpage
%\bibliographystyle{IEEEtran}

%\balance
%\bibliography{BibDataBase}

\begin{thebibliography}{10}
\providecommand{\url}[1]{#1}
\csname url@samestyle\endcsname
\providecommand{\newblock}{\relax}
\providecommand{\bibinfo}[2]{#2}
\providecommand{\BIBentrySTDinterwordspacing}{\spaceskip=0pt\relax}
\providecommand{\BIBentryALTinterwordstretchfactor}{4}
\providecommand{\BIBentryALTinterwordspacing}{\spaceskip=\fontdimen2\font plus
\BIBentryALTinterwordstretchfactor\fontdimen3\font minus
  \fontdimen4\font\relax}
\providecommand{\BIBforeignlanguage}[2]{{%
\expandafter\ifx\csname l@#1\endcsname\relax
\typeout{** WARNING: IEEEtran.bst: No hyphenation pattern has been}%
\typeout{** loaded for the language `#1'. Using the pattern for}%
\typeout{** the default language instead.}%
\else
\language=\csname l@#1\endcsname
\fi
#2}}
\providecommand{\BIBdecl}{\relax}
\BIBdecl

\bibitem{hua14}
J.~Huang, X.~Liang, X.~Qin, P.~Xie, and C.~Xie, ``Scale-{RS}: An efficient
  scaling scheme for {RS}-coded storage clusters,'' \emph{IEEE Trans. Parallel
  and Distributed Systems}, vol.~26, no.~6, pp. 1704--1717, Jun. 2015.

\bibitem{hua07}
C.~Huang, M.~Chen, and J.~Li, ``Pyramid codes: Flexible schemes to trade space
  for access efficiency in reliable data storage systems,'' in \emph{Proc. IEEE
  Int. Symp. Network Comput. and Appl. (NCA)}, Cambridge, MA, Jul. 2007.

\bibitem{hua12}
C.~Huang, H.~Simitci, Y.~Xu, A.~Ogus, B.~Calder, P.~Gopalan, J.~Li, and
  S.~Yekhanin, ``Erasure coding in {W}indows {A}zure storage,'' in \emph{Proc.
  USENIX Annual Technical Conf.}, Boston, MA, Jun. 2012.

\bibitem{sat13}
M.~Sathiamoorthy, M.~Asteris, D.~Papailiopoulos, A.~G. Dimakis, R.~Vadali,
  S.~Chen, and D.~Borthakur, ``{XOR}ing elephants: Novel erasure codes for big
  data,'' in \emph{Proc. 39th Very Large Data Bases Endowment (VLDB)}, Trento,
  Italy, Aug. 2013.

\bibitem{Pap14}
D.~S. Papailiopoulos and A.~G. Dimakis, ``Locally repairable codes,''
  \emph{IEEE Trans. Inf. Theory}, vol.~60, no.~10, pp. 5843--5855, Oct. 2014.

\bibitem{dim10}
A.~G. Dimakis, P.~B. Godfrey, Y.~Wu, M.~J. Wainwright, and K.~Ramchandran,
  ``Network coding for distributed storage systems,'' \emph{IEEE Trans. Inf.
  Theory}, vol.~56, no.~9, pp. 4539--4551, Sep. 2010.

\bibitem{Ras11}
K.~V. Rashmi, N.~B. Shah, and P.~V. Kumar, ``Optimal exact-regenerating codes
  for distributed storage at the {MSR} and {MBR} points via a product-matrix
  construction,'' \emph{IEEE Trans. Inf. Theory}, vol.~57, no.~8, pp.
  5227--5239, Aug. 2011.

\bibitem{Rou10}
S.~{El Rouayheb} and K.~Ramchandran, ``Fractional repetition codes for repair
  in distributed storage systems,'' in \emph{Proc. 48th Annual Allerton Conf.
  Commun., Control, and Comput.}, Monticello, IL, Sep./Oct. 2010.

\bibitem{wan14}
Y.~Wang, X.~Yin, and X.~Wang, ``{MDR} codes: A new class of {RAID}-6 codes with
  optimal rebuilding and encoding,'' \emph{IEEE J. Sel. Areas Commun.},
  vol.~32, no.~5, pp. 1008--1018, May 2014.

\bibitem{tam13}
I.~Tamo, Z.~Wang, and J.~Bruck, ``Zigzag codes: {MDS} array codes with optimal
  rebuilding,'' \emph{IEEE Trans. Inf. Theory}, vol.~59, no.~3, pp. 1597--1616,
  Mar. 2013.

\bibitem{Cad11}
V.~R. Cadambe, C.~Huang, J.~Li, and S.~Mehrotra, ``Polynomial length {MDS}
  codes with optimal repair in distributed storage,'' in \emph{Proc. 45th
  Asilomar Conf. Signals, Syst. and Comput. (ASILOMAR)}, Pacific Grove, CA,
  Nov. 2011.

\bibitem{Aga15}
G.~K. Agarwal, B.~Sasidharan, and P.~V. Kumar, ``An alternate construction of
  an access-optimal regenerating code with optimal sub-packetization level,''
  in \emph{Proc. 21st Nat. Conf. Commun. (NCC)}, Mumbai, India, Feb. 2015.

\bibitem{Li15}
J.~Li, X.~Tang, and U.~Parampalli, ``A framework of constructions of minimal
  storage regenerating codes with the optimal access/update property,''
  \emph{IEEE Trans. Inf. Theory}, vol.~61, no.~4, pp. 1920--1932, Apr. 2015.

\bibitem{Wan16}
Z.~Wang, I.~Tamo, and J.~Bruck, ``Explicit minimum storage regenerating
  codes,'' \emph{IEEE Trans. Inf. Theory}, vol.~62, no.~8, pp. 4466--4480, Aug.
  2016.

\bibitem{Sas16}
\BIBentryALTinterwordspacing
B.~Sasidharan, M.~Vajha, and P.~V. Kumar, ``An explicit, coupled-layer
  construction of a high-rate {MSR} code with low sub-packetization level,
  small field size and all-node repair,'' Sep. 2016, arXiv: 1607.07335v3.
  [Online]. Available: \url{https://arxiv.org/abs/1607.07335}
\BIBentrySTDinterwordspacing

\bibitem{Ye17}
M.~Ye and A.~Barg, ``Explicit constructions of optimal-access {MDS} codes with
  nearly optimal sub-packetization,'' \emph{IEEE Trans. Inf. Theory}, vol.~63,
  no.~10, pp. 6307--6317, Oct. 2017.

\bibitem{Li17}
J.~Li, X.~Tang, and C.~Tian, ``A generic transformation for optimal repair
  bandwidth and rebuilding access in {MDS} codes,'' in \emph{Proc. IEEE Int.
  Symp. Inf. Theory (ISIT)}, Aachen, Germany, Jun. 2017.

\bibitem{Rav17}
N.~Raviv, N.~Silberstein, and T.~Etzion, ``Constructions of high-rate minimum
  storage regenerating codes over small fields,'' \emph{IEEE Trans. Inf.
  Theory}, vol.~63, no.~4, pp. 2015--2038, Apr. 2017.

\bibitem{Ras17}
K.~V. Rashmi, N.~B. Shah, and K.~Ramchandran, ``A piggybacking design framework
  for read-and download-efficient distributed storage codes,'' \emph{IEEE
  Trans. Inf. Theory}, vol.~63, no.~9, pp. 5802--5820, Sep. 2017.

\bibitem{Hou16}
H.~Hou, K.~W. Shum, M.~Chen, and H.~Li, ``{BASIC} codes: Low-complexity
  regenerating codes for distributed storage systems,'' \emph{IEEE Trans. Inf.
  Theory}, vol.~62, no.~6, pp. 3053--3069, Jun. 2016.

\bibitem{Hou17}
H.~Hou, P.~P.~C. Lee, Y.~S. Han, and Y.~Hu, ``Triple-fault-tolerant binary
  {MDS} array codes with asymptotically optimal repair,'' in \emph{Proc. IEEE
  Int. Symp. Inf. Theory (ISIT)}, Aachen, Germany, Jun. 2017.

\bibitem{kum15a}
S.~Kumar, A.~{Graell i Amat}, I.~Andriyanova, and F.~Br\"annstr\"om, ``A family
  of erasure correcting codes with low repair bandwidth and low repair
  complexity,'' in \emph{Proc. IEEE Global Commun. Conf. (GLOBECOM)}, San
  Diego, CA, Dec. 2015.

\bibitem{Yua16}
S.~Yuan and Q.~Huang, ``Generalized piggybacking codes for distributed storage
  systems,'' in \emph{Proc. IEEE Global Commun. Conf. (GLOBECOM)}, Washington,
  DC, Dec. 2016.

\bibitem{Raw17}
A.~S. Rawat, I.~Tamo, V.~Guruswami, and K.~Efremenko, ``{MDS} code
  constructions with small sub-packetization and near-optimal repair
  bandwidth,'' \emph{IEEE Trans. Inf. Theory}, to appear.

\bibitem{Bla95}
M.~Blaum, J.~Brady, J.~Bruck, and J.~Menon, ``{EVENODD}: An efficient scheme
  for tolerating double disk failures in {RAID} architectures,'' \emph{IEEE
  Trans. Comput.}, vol.~44, no.~2, pp. 192--202, Feb. 1995.

\bibitem{Kar63}
A.~Karatsuba and Y.~Ofman, ``Multiplication of multidigit numbers on
  automata,'' \emph{Sov. Phys. Doklady}, vol.~7, no.~7, pp. 595--596, Jan.
  1963.

\bibitem{Wei06}
\BIBentryALTinterwordspacing
A.~Weimerskirch and C.~Paar, ``Generalizations of the {K}aratsuba algorithm for
  efficient implementations,'' Jul. 2006. [Online]. Available:
  \url{https://eprint.iacr.org/2006/224.pdf}
\BIBentrySTDinterwordspacing

\bibitem{Pol71}
J.~M. Pollard, ``The fast {F}ourier transform in a finite field,'' \emph{Math.
  Comput.}, vol.~25, no. 114, pp. 365--374, Apr. 1971.

\bibitem{Cra01}
R.~Crandall and C.~Pomerance, \emph{Prime Numbers: A Computational
  Perspective}.\hskip 1em plus 0.5em minus 0.4em\relax Springer, 2001.

\bibitem{Gao10}
S.~Gao and T.~Mateer, ``Additive fast {F}ourier transforms over finite
  fields,'' \emph{IEEE Trans. Inf. Theory}, vol.~56, no.~12, pp. 6265--6272,
  Dec. 2010.

\end{thebibliography}

% Generated by IEEEtran.bst, version: 1.14 (2015/08/26)

\end{document}